\pdfoutput=1

\documentclass{scrartcl}

\KOMAoptions{fontsize=10}

\usepackage{ownstyle}
\title
{Initial Semantics for higher-order typed syntax in \textsf{Coq}}

\author
{Benedikt Ahrens\\{\small Universit\'e Nice -- Sophia Antipolis, France}\\ {\small \href{mailto:ahrens@unice.fr}{ahrens@unice.fr}}
\and Julianna Zsid\'o\\{\small Universit\'e  Montpellier II, France}\\ {\small \href{mailto:jzsido@univ-montp2.fr}{jzsido@univ-montp2.fr}}}

\date{}

\usepackage{listings}

\begin{document}

\maketitle

\begin{abstract}
Initial Semantics aims at characterizing the syntax associated to a signature as the initial object of 
some category. 
We present an initial semantics
result for typed syntax with variable binding together with its formalization in the \textsf{Coq} proof assistant.
The main theorem was first proved on paper in the second author's PhD thesis in 2010, 
and verified formally shortly afterwards. %

To a simply--typed binding signature $S$ over a fixed set $T$ of object types we associate a category called the 
\emph{category of representations of $S$}. 
We show that this category has an \emph{initial object $\Sigma(S)$}, i.e.\ an object $\Sigma(S)$ 
from which there is precisely one morphism $i_R:\Sigma(S) \to R$ to any object $R$ of this category.
From its construction it will be clear that the object $\Sigma(S)$ merits the name \emph{abstract syntax associated to $S$}:
it is given by an inductive %
set 
--- parametrized by a set of free variables and dependent on object types --- 
the type of whose constructors are each given by the arities of the signature $S$.

Our theorem is implemented and proved correct in the proof assistant  \textsf{Coq} through heavy use of dependent types. 
The approach through monads gives rise to an implementation of syntax where both terms and variables are \emph{intrinsically typed}, 
i.e.\ where the object types are reflected in the meta--level types. %
Terms are implemented as a \textsf{Coq} data type --- \textsf{Coq} types play the role of sets --- dependent on an 
object type as well as on a type family of free variables.

This article is to be seen as a research article rather than about the formalization of a classical mathematical result.
The nature of our theorem -- involving lengthy, technical proofs %
 and complicated algebraic structures -- makes it particularly interesting 
for formal verification. Our goal is to promote the use of computer theorem provers as research tools, and, accordingly, a new way of publishing 
mathematical results: a parallel description of a theorem and its formalization should allow the verification of correct 
transcription of definitions and statements into the proof assistant, and %
straightforward but technical proofs should be well--hidden in a digital library.
We argue that \textsf{Coq}'s rich type theory, combined with its various features such as implicit arguments, %
allows a particularly readable formalization and is hence well--suited for communicating mathematics.
\end{abstract}


\newpage \tableofcontents \newpage

\section{Introduction}

Computer theorem proving is a subject of active research, and provers are under heavy 
development, evolving rapidly. 
However, we believe that the provers at hand --- and in particular, our favourite 
prover \textsf{Coq} \cite{coq} --- have reached a state where they are well usable as a research tool.
Instead of benchmarking it with one of the classical mathematical results, as is done e.g.\ in 
Wiedijk's list ``Formalizing 100 theorems'' \footnote{\url{http://www.cs.ru.nl/~freek/100/index.html}} 
(cf.\ also \cite{wiedijk_formalproof}), we use \textsf{Coq} to prove a recent theorem
about typed abstract syntax with variable binding
\footnote{We use the term ``higher--order'' synonymous to ``with variable binding''. 
          The term is also used in the expression ``Higher--Order Abstract Syntax'', where it refers to the 
	    way in which variable binding is modeled, e.g.\ as in $lam:(T \to T) \to T$.
          We do \emph{not} model variable binding in this way.
}.
Through the use of \textsf{Coq} features such as \emph{implicit arguments, coercions} and 
\emph{overloading through type classes}
the formal text remains close to its informal counterpart, thus easing the verification of correct
transcription of definitions and statements into the formal language.

Category--theoretic concepts have been introduced to computer science, more specifically to programming,
in order to give mathematical structure to programs, e.g.\ by Wadler \cite{DBLP:conf/afp/Wadler95}.
This development culminates in the programming language \textsc{Haskell}, whose basic programming idioms are indeed 
category--theoretic notions. In particular, the notion of \emph{monad}, which we also use extensively, has 
a prominent r\^ole in \textsc{Haskell}.

 In his PhD thesis, Vene \cite{vene} studies different classes of \emph{recursive functions}
and characterizes them as morphisms in some category.



All these examples concern category theoretic concepts which can be found \emph{within} the programming language,
i.e.\ on the object level. In this paper, however, category theory is used on the \emph{meta level} in order to give a definition of the 
programming language associated to a signature. 

Indeed, our goal is to characterize the set of terms of a language given by a \emph{typed binding signature} via a \emph{universal property},
and give a \emph{category--theoretic justification} for the \emph{recursion principle} it is equipped with.

A universal property characterizes its associated object --- if it exists --- up to a unique isomorphism, for a suitable notion of morphism.
 Universal properties are ubiquitous in mathematics, and fundamental concepts such as the cartesian product of two sets,
  the free group associated to a set or the field of quotients associated to an integral domain can be defined as 
  objects verifying a suitable universal property.

The universal property we use to characterize syntax is \emph{initiality} (cf.\ Def.\ \ref{def:initial_object}):
given a signature $S$, we construct a category in which the syntax $\Sigma(S)$ associated to $S$ is initial,
thus characterizing $\Sigma(S)$ up to isomorphism.

This is precisely what the expression ``Initial Semantics'' stands for: the objects of this category can be thought of as ``semantics'' of
$S$, and the syntax $\Sigma(S)$ is the initial such semantics
\footnote{We use the word ``semantics'' with two different meanings. Accompanied by the word ``initial'', i.e. in the expression ``initial semantics'',
    it refers to the syntax associated to some signature $S$ being the initial ``model'' or ``semantics'', in a category
          of ``semantics of $S$''. 
       The word ``semantics'' by itself signifies a relation on terms, usually a reduction relation, e.g.\ beta reduction.}.


In this paper, category--theoretic concepts appear in two places:
firstly, as explained above, we characterize the syntax $\Sigma (S)$
associated to a signature $S$ as the initial object of some category. 
Secondly, the objects of said category are built from \emph{monads} (cf.\ Def.\ \ref{def:monad}) over the category of (families of) sets.
Indeed, we consider an untyped programming language to be given by such a monad, i.e.\ a map which 
associates to any set $V$ a set of terms with free variables in $V$, together with some extra structure 
(cf.\ Ex.\ \ref{ex:ulc_monad}).
For simply--typed syntax over a set $T$ of types, we regard families of sets, indexed by $T$, 
rather than just sets, cf.\ Ex.\ \ref{ex:slc_monad}.

We consider the syntax $\Sigma(S)$ to be given as an inductive family of sets, para\-me\-trized by free variables 
and indexed by the set of object types. Initial Semantics can hence also be seen as the study of 
a restricted class of inductive data types.

In Subsec.\ \ref{subsec:informal_intro} we introduce initiality using a particularly simple inductive set --- the natural numbers
--- and outline its generalization to abstract syntax as a para\-me\-trized and dependent 
inductive type.
In Subsec.\ \ref{subsec:overview} we give a technical overview of the paper.
In Subsec.\ \ref{sec:rel_work} we give an overview over various initial semantics results.

The complete \textsf{Coq} code can be obtained from the first author's web page
\footnote{\url{http://math.unice.fr/~ahrens}}.

\subsection{Inductive Types, Categorically}\label{subsec:informal_intro}

Initial Semantics has its origins in the \emph{Initial Algebras} as studied by Goguen et al.\ \cite{gtww}.
It can be considered as a category--theoretic treatment of recursion and induction.
A prominent example is given by the Peano axioms:
consider the category $\mathcal{N}$ an object of which is a triple $(X,Z,S)$ of a set $X$ 
together with a constant $Z \in X$ and a unary operation
$S:X\to X$. 
A morphism to another such object $(X',Z', S')$ is a map $f : X \to X'$
such that 
\begin{equation} f(Z) = Z' \quad \text{ and } \quad f\circ S = S' \circ f \enspace . \label{eq:nat_mor} \end{equation}
This category has an initial object $(\mathbb{N}, \Zero, \Succ)$ given by the natural numbers $\mathbb{N}$
equipped with the constant $\Zero = 0$ and the successor function $\Succ$.
Initiality of $\mathbb{N}$ gives a way to define \emph{iterative} functions \cite{vene}
from $\mathbb{N}$ to any set $X$ by equipping $X$ with a constant $Z\in X$ and 
a unary map $S:X\to X$,
i.e.\ making the set $X$ the carrier of an object $(X,Z,S)\in\mathcal{N}$. 

Using the preceding example, we now informally introduce some vocabulary 
which is used (and properly defined) later.
For specifying a syntax, an \emph{arity} indicates the number of arguments of a \emph{constructor}.
The arities of $Z$ and $S$ are $0$ and $1$, respectively.
A \emph{representation} of an arity $n$ in a set $X$ is then given by an $n$--ary operation on $X$.
A \emph{signature} is a family -- indexed by some arbitrary set $J$ -- of arities.
A representation of a signature is given by a set $X$ and a representation of each arity of $S$ in $X$.
The signature $\mathcal{N}$ of the preceding example is given by 
\[ \mathcal{N} := \{z \mapsto 0 \enspace ,\quad s \mapsto 1\} \enspace , \] 
and a representation
of this signature is any triple $(X,S,Z)$ as above.

\subsubsection*{Adding variables}
When considering syntax \emph{with variable binding}, the set of terms is indexed by a set of variables 
whose elements may appear 
freely in those terms. 
\begin{example}\label{ex:ulc_def}
 As an example, consider the following inductive set $\LC:\Set\to\Set$ of terms of the untyped lambda calculus:
\begin{align*} 
\LC (V) ::=\quad &\Var : V \to \LC(V) \\ 
        {}\mid{} &\Abs : \LC (V^{*}) \to \LC (V) \\ 
        {}\mid{} &\App : \LC(V) \to \LC(V) \to \LC(V) \enspace ,
\end{align*}
where $V^{*} := V + \lbrace {*} \rbrace$ 
is the set $V$ enriched with a new distinguished variable --- the variable which is
bound by the $\Abs$ constructor (cf.\ Sec.\ \ref{sec:deriv}).
We continue this example in the course of the paper (cf.\ Ex.\ \ref{ex:ulc_monad}, 
\ref{ex:ulc_taut_mod}, \ref{ex:ulc_const_mod} , \ref{ex:ulc_mod_mor}, \ref{ex:ulc_rep}, \ref{ex:ulc_rep_mor}).
\end{example}
In this case arities need to carry information about the binding behaviour of the constructor they are associated to.
One way to define such arities is using lists of natural numbers. 
The length of a list then indicates the number of arguments of
the constructor, %
and the $i$-th entry denotes the number 
of variables that the constructor binds in the $i$-th argument.
The signature $\mathcal{LC}$ of $\LC$ is given by 
\[ \mathcal{LC} := \{ \app \mapsto [0,0]\enspace , \quad \abs \mapsto [1] \} \enspace . \]

Representations in sets are not adequate any more for such a syntax;
instead we should represent the signature $\mathcal{LC}$ in objects
with the same type as $\LC$, i.e.\ in maps $F : \Set \to \Set$ associating a set $F(V)$ 
to any given set $V$ ``of variables''.
Accordingly, a representation of an arity now is not simply an $n$--ary operation, but 
a family of maps, indexed by the set $V$ of variables. 
Indeed, a representation of, e.g.\ the arity $\abs$ of $\mathcal{LC}$, in a suitable map $F:\Set\to\Set$, 
should have the same type as the constructor $\Abs$, that is, 
\[  \abs^F (V) : F(V^*) \to F(V) \enspace . \]

\subsubsection*{Interlude on monads}
Instead of maps $F:\Set\to\Set$ as in the preceding paragraph, we consider in fact
\emph{monads} on the category $\Set$ of sets. 
Monads are such maps %
equipped with some extra structure, which we explain by the example of the untyped
lambda calculus.
The map $ V\mapsto \LC(V)$ comes with a (capture--avoiding) substitution operation:
let $V$ and $W$ be two sets (of variables) and $f$ be a map $f : V\to \LC(W)$.
Given a lambda term $t\in \LC(V)$, we can replace each free variable $v\in V$
in $t$ by its image under $f$, yielding a term $t' \in \LC(W)$.
Furthermore we consider the constructor $\Var_V$ as a ``variable--as--term'' map, indexed by a set of variables $V$, 
\[\Var_V : V\to \LC(V) \enspace . \]
There is a well--known algebraic structure which captures those two operations and their properties:
 substitution and variable--as--term map 
turn $\LC$ into a \emph{monad} (Def.\ \ref{def:monad}) on the category of sets, an observation first made by Altenkirch and Reus \cite{alt_reus}.
We expand on this in Ex.\ \ref{ex:ulc_monad}.


The monad structure of $\LC$ should be compatible in a suitable sense with the constructors $\Abs$ and $\App$ of $\LC$. 
One mathematical structure which would express such a compatibility is that of a \emph{monad morphism}.
This fails in 2 ways:

firstly, it is unclear how to equip the domain map $V\mapsto \LC(V)\times\LC(V)$ of $\App$ with a monad structure.

Secondly, while the domain of the constructor $\Abs$, the map $\LC^* : V \mapsto \LC(V^*)$, inherits a monad structure from $\LC$ (cf.\ Ex.\ \ref{ex:derived_monad}),
the constructor $\Abs$ does not verify the properties of a morphism of monads (cf.\ Ex.\ \ref{ex:abs_not_monadic} and \cite{DBLP:conf/wollic/HirschowitzM07}).

As a remedy, Hirschowitz and Maggesi \cite{DBLP:conf/wollic/HirschowitzM07} consider \emph{modules over a monad} (cf.\ Def.\ \ref{def:module}), 
which generalize monadic substitution, and suitable morphisms of modules.
Indeed, the maps $\LC : V\mapsto \LC(V)$ and $\LC^* : V\mapsto \LC(V^*)$ are 
the underlying maps of such modules (cf.\ Ex.\ \ref{ex:ulc_taut_mod}, \ref{ex:ulc_const_mod}), and
the constructors $\Abs$ and $\App$ are morphisms of modules (cf.\ Ex.\ \ref{ex:ulc_mod_mor}).


\subsubsection*{Typed syntax} Typed syntax exists with varying complexity, ranging from simply--typed
syntax to syntax with dependent types, kinds, polymorphism, etc.
By simply--typed syntax we mean a non--polymorphic %
typed syntax where the set of types is independent 
from the set of terms, 
i.e.\ one has a fixed set of types, the elements of which are used to type variables and terms. 
A simply--typed syntax does not allow type constructors in its associated signatures, only (typed) term constructors.
In more sophisticated type systems types may depend on terms, leading to 
more complex definitions of arities and signatures.

This work is only concerned with simply--typed languages, such as the simply--typed
lambda calculus and PCF.
For such a simply--typed syntax, we first fix a set $T$ of (object) types.
Variables then are equipped with a type $t\in T$, i.e.\ instead of \emph{one} set of variables
we consider a family $(V_t)_{t \in T}$ of sets of variables, where $V_t$ is the set of variables of type $t$.
Similarly the terms of a simply--typed syntax come as a family of sets, indexed by the (object) types.
As an example we consider the simply--typed lambda calculus $\SLC$:  %

\begin{example}\label{ex:slc_def}
 Let $\mathcal{T} ::= * \mid \mathcal{T} \Rightarrow \mathcal{T}$ be the set of types of the simply--typed lambda calculus. 
 For each family $V : \mathcal{T} \to \Set$ of sets and $t\in \mathcal{T}$ we denote by $V_t := V(t)$ 
the set associated to object type $t$.
The set of simply--typed lambda terms with free variables in the family of sets $V$ is given by the following inductive declaration:
\begin{align*} 
\SLC (V) : \mathcal{T}\to\Set\enspace ::=\quad &\Var : \forall t, ~V_t \to \SLC(V)_t \\ 
        {}\mid{} &\Abs : \forall s~t, ~\SLC (V^{*s})_t \to \SLC (V)_{(s\Rightarrow t)} \\ 
        {}\mid{} &\App : \forall s~t, ~\SLC(V)_{(s\Rightarrow t)} \to \SLC(V)_s\to \SLC(V)_t \enspace ,
\end{align*}
where $V^{*s} := V + \lbrace {*s} \rbrace$ is obtained by enriching the family $V$ with a new distinguished 
variable of type $s\in \mathcal{T}$ --- the variable which is
bound by the constructor $\Abs{(s,t)}$. %
The variables $s$ and $t$ range over the set $\mathcal{T}$ of types.
The signature describing the simply--typed lambda calculus is given in Ex.\ \ref{ex:slc_arities}. 
The preceding paragraph about monads and modules applies to the simply--typed lambda calculus when replacing sets by 
families of sets indexed by $\mathcal{T}$:
the simply--typed lambda calculus can be given the structure of a monad (cf.\ Ex.\ \ref{ex:slc_monad})
\[ \SLC : \TS{\mathcal{T}} \to \TS{\mathcal{T}} \enspace  \]
over the category of families of sets indexed by $\mathcal{T}$ (Def.\ \ref{def:TST}).
The constructors of $\SLC$ are morphisms of modules (cf.\ Ex.\ \ref{ex:slc_modules}, \ref{ex:slc_mod_mor}).

\end{example}

\subsection{Overview of the paper}\label{subsec:overview}

We present an initial semantics result and its formalization for typed higher--order syntax with types. %
The term ``higher--order'' refers to the fact that the syntax allows for variable binding in terms.
Our types are, more specifically, \emph{simple types}, e.g.\ there is no binding on the level of types.

Our theorem is not the first of its kind, cf. Sec.\ \ref{sec:rel_work} for related work.
It is, however, the only one which is based on monads and modules and is fully implemented in a proof assistant.

In order to account for types, our basic category of interest is the category $[T,\Set]$ of families of sets indexed by a set $T$.
Its objects will also be called ``typed sets'' 
Our monads are monads over $[T,\Set]$.

The notion of \emph{module over a monad} \cite{DBLP:conf/wollic/HirschowitzM07} generalizes monadic substitution: 
a module is a functor with a substitution map.
Morphisms of modules are natural transformations which are compatible with the module substitution.

We interpret the syntax associated to a signature $S$ as an initial object 
in the category of so--called \emph{representations of $S$}. An object of this 
category is a monad over typed sets equipped with a morphism of modules for each arity of $S$. 
A morphism of representations is a morphism between the underlying monads which is compatible with the morphisms of modules.
For the initial representation these module morphisms are given by the \emph{constructors} of the syntax, 
and the property of being a module morphism captures their compatibility with substitution.

Our theorem is implemented in the proof assistant \textsf{Coq} \cite{coq}. 
This implementation can be seen as a formal proof of a mathematical theorem in a constructive setting, 
and as such delivers confidence in the correctness of the theorem.

Perhaps more importantly, the theorem translates to an implementation of syntax using exclusively \emph{intrinsic typing}, 
a style of implementation that has been advertised by Benton et al. \cite{dep_syn}. 
Here typing is not done by a typing judgement, given by, say, an inductive predicate. 
Instead it relies on type parameters, i.e.\ on dependent types, in the meta--language. The technique and its benefits are discussed in \cite{dep_syn}.

\subsection{Related Work} \label{sec:rel_work}
The theorem we present was first proved in Zsid\'o's PhD thesis \cite{ju_phd}. 
It is a generalization of the work by Hirschowitz and Maggesi on untyped syntax \cite{DBLP:journals/iandc/HirschowitzM10} 
 based on the notion of monads and modules over monads.
Monads were identified by 
Altenkirch and Reus \cite{alt_reus} as a convenient categorical device to talk about substitution. 

\paragraph*{Initial semantics}

For untyped first-order syntax the notion of \emph{initial algebra} was coined by Goguen et al.\ \cite{gtww} in the 1970s. 
%


Initial semantics has then been extended to account for additional features, 
as illustrated by the following scheme: %

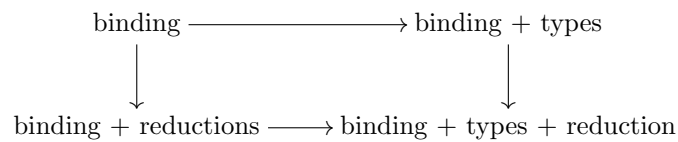
\[ 
   \begin{xy}
    \xymatrix{ \text{binding} \ar[r] \ar[d] &  \text{binding + types} \ar[d] \\
               \text{binding + reductions} \ar[r] &  \text{binding + types + reduction}
}
   \end{xy}
\]

Another criterion to classify initiality results is the way in which variable binding is modeled. 
Frequently used for representing binding are the following techniques:

\begin{enumerate}
 \item Nominal syntax using named abstraction, \label{list:nominal}
  \item Higher--Order Abstract Syntax (HOAS), e.g.\ $lam : (T \to T) \to T$ and its \emph{weak} variant, e.g.\ $lam : (var \to T) \to T$ \label{list:hoas} and
 \item Nested Datatypes as introduced in \cite{BirdMeertens98:Nested}. \label{list:brujin}
\end{enumerate}

\noindent
Initial semantics for untyped syntax were presented by 
Gabbay and Pitts \cite[(\ref{list:nominal})]{gabbay_pitts99} ,
Hofmann \cite[(\ref{list:hoas})]{hofmann}
and Fiore et al.\ \cite[(\ref{list:brujin})]{fpt}. The numbers given in parentheses correspond to the way variable binding is modeled, according to 
the list given above.
Hirschowitz and Maggesi \cite[(\ref{list:brujin})]{DBLP:conf/wollic/HirschowitzM07} prove an initiality result for arbitrary untyped syntax based on the notion of 
\emph{monads}. %

The extension to simply--typed syntax was done, for the HOAS approach, by Miculan and Scagnetto \cite[(\ref{list:hoas})]{DBLP:conf/ppdp/MiculanS03}.

Fiore et al.'s approach was generalized to encompass the simply--typed lambda calulus in \cite{fio02}, and detailed for general simply--typed syntax
in Zsid\'o's PhD thesis \cite{ju_phd}.

There, she also generalized Hirschowitz and Maggesi's approach \cite{DBLP:conf/wollic/HirschowitzM07} to simply--typed syntax. 
It is this result and its formalization in \textsf{Coq} that the present article is about.

Both lines of work, Hirschowitz and Maggesi's and Fiore et al.'s, are deeply connected. Zsid\'o \cite{ju_phd} made this connection precise,
by establishing an adjunction between the resp.\ categories under consideration.

\emph{Semantic aspects} were integrated in initiality results by several people. 

Hirschowitz and Maggesi \cite{DBLP:conf/wollic/HirschowitzM07} characterize the terms of the lambda calculus 
modulo beta and eta reduction as an initial object in some category.

Another idea mentioned in \cite{DBLP:conf/wollic/HirschowitzM07} is to consider not sets of terms, quotiented by 
reduction relations, but sets equipped with a preorder. This idea is being pursued by the first author.

Fiore and Hur \cite{DBLP:conf/icalp/FioreH07} extended Fiore et al.'s approach to ``second--order universal algebras''. 
In particular, Hur's PhD thesis \cite{hur_phd} is dedicated to this extension.

While the present paper does not treat semantic aspects, one of the goals is to 
set up and formalize the techniques which will be necessary for understanding
semantic aspects in the simply typed case.


\paragraph*{Implementation of syntax}

The implementation and formalization of syntax has been studied by a variety of people. 
The \popl challenge \cite{poplmark} is a benchmark which aims to evaluate readability and provability when using different techniques of variable binding. 
The technique we use, called \emph{Nested Abstract Syntax}, is used in a partial solution by Hirschowitz and Maggesi \cite{NAS}, 
but was proposed earlier by others, e.\ g.\ \cite{BirdMeertens98:Nested, alt_reus}. 
The use of \emph{intrinsic typing} by dependent types of the meta--language was advertised in \cite{dep_syn}.

During our work we became aware of Capretta and Felty's framework for reasoning about programming languages \cite{HOAU}.
They implement a tool --- also in the \textsf{Coq} proof assistant --- which, given a signature, provides 
the associated abstract syntax as a data type dependent on the object types, hence intrinsically typed as well. 
Their data type of terms does not, however, depend on the set of free variables of those terms. Variables are encoded with de Brujin indices.
There are two different constructors for free and bound variables which serve to control the binding behaviour of object level constructors.
In our theorem, there is only one constructor for (free) variables, and binding a variable is done by
removing it from the set of free variables. 

Capretta and Felty then add a layer to translate those terms into syntax using named abstraction, and provide suitable induction and recursion principles. 
Their tool may hence serve as a practical framework for reasoning about programming languages.
Our implementation remains on the theoretical side by not providing named syntax and exhibiting the category--theoretic properties of abstract syntax.

\subsection*{Synopsis}

In the second section we give a very brief description of \textsf{Coq}, the theorem prover we use for the formalization. 
Afterwards we explain how we deal with the problem of formalizing algebraic structures.

\noindent
The third section presents categorical concepts and their formalization. 
We state the definition of category, initial object of a category, \emph{monad} (as Kleisli structure) and \emph{module over a monad} as 
well as their resp.\ morphisms. Some constructions on monads and modules are explained, which will be of importance in what follows.

\noindent
The fourth section introduces the notions of \emph{arity}, \emph{signature} and \emph{representations} of signatures in suitable monads. 
The category of representations of a given signature is defined. The main theorem \ref{nice_thm} states that this category has an initial object.

\noindent
In the fifth part the formal construction of said initial object is explained.

\noindent
Some conclusions and future work are stated in the last section.

\section{Preliminaries}

\subsection{About the proof assistant \textsf{Coq}}

The proof assistant \textsf{Coq} \cite{coq} is an implementation of the 
\emph{Calculus of Inductive Constructions (CIC)} which itself is a constructive \emph{type theory}. 
Bertot and Casteran's book \emph{Coq'Art} \cite{L:BC04} gives a comprehensive introduction to \textsf{Coq}. 
The \textsf{Coq} web page \cite{coq} carries links to more howtos and specialised tutorials.
In \textsf{Coq} a typing judgment is written \lstinline!t : T!, meaning that $t$ is a term of type $T$. 
Function application is simply denoted by a blank, i.e.\ we write \lstinline!f x! for $f(x)$.

The CIC also treats propositions as types via the \emph{Curry--Howard isomorphism}, hence a proof of a proposition $P$ is in fact a term of type $P$.  
In the proof assistant \textsf{Coq} a user hence proves a proposition \lstinline!P! by providing a term \lstinline!p! of type \lstinline!P!. 
\textsf{Coq} checks the validity of the proof \lstinline!p! by verifying whether \lstinline!p : P!.

\textsf{Coq} comes with extensive support to \emph{interactively build} the proof terms of a given proposition.
 In \emph{proof mode} so-called \emph{tactics} help the user to reduce the proposition they want to prove -- the \emph{goal} -- into one or more simpler subgoals, 
until reaching trivial subgoals which can be solved directly.

Particular concepts of \textsf{Coq} such as records and type classes, setoids, implicit arguments and coercions are explained 
in a call-by-need fashion in the course of the paper.
One important feature is the \lstinline!Section! mechanism (cf.\ also the \textsf{Coq} manual \cite{CoqManualV83}). 
Parameters and hypotheses declared in a section automatically get 
discharged when closing the section. Constants of the section then become functions,
depending on an argument of the type of the parameter they mentioned.
When necessary, we will either give a slightly modified, fully discharged version of a statement, or mention 
the section parameters in the text.

\subsection{How to formalize algebraic structures}\label{sec:alg_structure}

The question of how to formalize algebraic structures is a subject of active research. 
We do not attempt to give an answer of any kind here. However, we need to choose from the existing solutions.

In \textsf{Coq} there are basically two possible answers: \emph{type classes} \cite{sozeau.Coq/classes/fctc}, as used 
by Spitters and v.~d.~Weegen \cite{DBLP:journals/mscs/SpittersW11} and \emph{records}, employed e.g.\ by Garillot et al.~\cite{packing}.

\textsf{Coq} records are implemented as an inductive data type with one constructor,
However, use of the vernacular command \lstinline!Record! (instead of plain \lstinline!Inductive!)
allows the optional automatic definition of the projection functions to the 
constructor arguments -- the ``fields'' of the record.
Additionally, one can declare those projections as \emph{coercions}, i.e.\ they can be inserted 
automatically by \textsf{Coq}, and left out in printing.
As an example for a coercion, it allows us to write \lstinline!c : C! for an object \lstinline!c!
of a category \lstinline!C!. Here the projection from the category type to the type of objects of 
a category is declared as a coercion (cf.\ Listing \ref{lst:Cat}). 
This is the formal counterpart to the convention introduced in the informal definition of categories 
in Def.\ \ref{def:category}.
Another example of coercion 
is given in the definition of monad (cf.\ Def.\ \ref{def:monad}), where it corresponds precisely to 
the there--mentioned \emph{abuse of notation}.

Type classes are implemented as records. Similarly to the difference between records and inductive types,
type classes are distinguished from records --- from a technical point of view --- only in 
that some meta--theoretic features are automatically enabled when 
declaring an algebraic structure as a class rather than a record.
For details we refer to Sozeau's article about the implementation of type classes \cite{sozeau.Coq/classes/fctc}
and Spitters and v.\ d.\ Weegen's work \cite{DBLP:journals/mscs/SpittersW11}.


Type classes differ from records in their usage, more specifically, in which data one declares as a \emph{parameter} of the
structure and which one declares as a \emph{field}.
The following example, borrowed from \cite{DBLP:journals/mscs/SpittersW11}, illustrates the different uses; we give 
two definitions of the algebraic structure of reflexive relation, one in terms of classes and one in terms of records:
\begin{lstlisting}
Class Reflexive {A : Type}{R : relation A} := 
   reflexive : forall a, R a a.

Record Reflexive := {
  carrier : Type ;
  car_rel : relation carrier ;
  rel_refl : forall a, car_rel a a }.
\end{lstlisting}


Our main interest in classes comes from the fact that by using classes many of the arguments of projections 
are automatically declared as \emph{implicit arguments}. This leads to easily readable code in that superfluous arguments which can be 
deduced by \textsf{Coq} do not have to be written down. Thus it corresponds precisely to 
the mathematical practice of not mentioning arguments (e.g.\ indices) which ``are clear from the context''. 
In particular, the structure argument of the projection, that is, the argument specifying the instance whose field 
we want to access, is implicit and deduced automatically by \textsf{Coq}. This mechanism allows for \emph{overloading},
a prime example being the implementation of setoids (cf.\ Sec.\ \ref{sec:setoids}) as a type class; 
in a term ``\lstinline!a == b!'' denoting setoidal equality, \textsf{Coq} automatically finds 
the correct setoid instance from the type of \lstinline!a! and \lstinline!b! 
\footnote{Beware! In case several instances of setoid have been declared on one and the same \textsf{Coq} type,
   the instance chosen by \textsf{Coq} might not be the one intended by the user. 
This is the main reason for Spitters and v.\ d.\ Weegen to restrict the 
     fields of type classes to \emph{propositions}.}.

We decide to define our algebraic structures in terms of type classes first, 
and bundle the class together with some of the class parameters in a record afterwards.
as is shown in the following example for the type class \lstinline!Cat_struct! (cf.\ Listing \ref{lst:cat}) 
and the bundling record \lstinline!Cat!.  
\begin{lstlisting}[caption={Bundling a type class into a record}, label={lst:Cat}]
Record Cat := {
     obj :> Type ;
     mor : obj -> obj -> Type ;
     cat_struct :> Cat_struct mor }.
\end{lstlisting}
In this code snippet the projections \lstinline!obj! and \lstinline!cat_struct! are defined as coercions, 
as explained at the beginning of this subsection, by using the notation ``\lstinline!:>!'' rather than just a colon.

The duplication of \textsf{Coq} definitions as classes \emph{and} records is a burden rather than a feature. We still proceed like this for the following reasons:

In our case the use of records is unavoidable since we want to have a \emph{\textsf{Coq} type} of categories, 
of functors between two given categories etc. 
This is necessary when categories, functors, etc.\ shall themselves be the objects or morphisms of some category,
as will be clear from Listing \ref{lst:cat}.
However, we profit from aforementioned features of type classes, notably automatic declaration of some arguments as \emph{implicit} and
the resulting overloading.

Apart from that, we do not employ any feature that makes the use of type classes comfortable --- 
such as maximally inserted arguments, operational classes, etc. --- 
since we usually work with the bundled versions.
Readers who want to know how to use type classes in \textsf{Coq} properly, 
should take a look at Spitters and v.\ d.\ Weegen's paper \cite{DBLP:journals/mscs/SpittersW11}.
They also employ the mentioned bundling of type classes in records whenever they need to build a 
\emph{category} of algebraic structures.
In the following we will only present the type class definition of each defined object.

\section{Categories, Monads \& Modules}

Mac Lane's book \cite{maclane} may serve as a reference for the following definitions, unless stated otherwise.
Note that we write ``$\comp{f}{g}$'' for the composite of morphisms $f : a \to b$ and $g : b \to c$ in any category, instead of
$g \circ f$.

\subsection{Categories}

\begin{definition}\label{def:category}
A \emph{category} $\C$ is given by
\begin{itemize}
 \item a collection -- which we will also call $\C$ -- of \emph{objects},
 \item for any two objects $c$ and $d$ of $\C$, a collection of \emph{morphisms}, written $\C(c,d)$,
 \item for any object $c$ of $\C$, a morphism $\id_c$ in $\C(c,c)$ and
 \item for any three objects $c, d, e$ of $\C$ a \emph{composition} operation
         \[ (\comp{\_}{\_})_{c,d,e} : \C(c,d) \times \C(d,e) \to \C(c,e) \]
\end{itemize}
such that the composition is associative and the morphisms of the form $\id_c$ for suitable objects $c$
are left and right neutral w.r.t.\ this composition
\footnote{We omit the ``object'' parameters from the composition operation, since those are deducible from 
  the morphisms we compose. This omission is done in our library as well, via \emph{implicit arguments} (cf.\ Sec.\ \ref{sec:alg_structure}).}:
\begin{align*}
  &\forall a~b~c~d : \C, \forall f : \C(a,b), g : \C(b,c), h : \C(d,e), \enspace \comp{f}{(\comp{g}{h})} = \comp{(\comp{f}{g})}{h} \\
  &\forall c~d : \C, \forall f : \C(c,d), \enspace \comp{f}{\id_d} = f \text{ and } \comp{\id_c}{f} = f \enspace .
\end{align*}
  We write $f:c\to d$ for a morphism $f$ of $\C(c,d)$.
\end{definition}

\begin{example}
  The category $\Set$ is the category of sets and, as morphisms from set $A$ to set $B$, the collection of total maps from $A$ to $B$, together with 
  the usual composition of maps.
\end{example}
\begin{definition}\label{def:TST}
  Let $T$ be a set. We denote by $\TS{T}$ the category whose objects are %
  collections of sets indexed by $T$. 
We also refer to such collections as \emph{type families} indexed by $T$,
since this is how we chose to implement them (cf.\ Sec.\ \ref{sec:coq_impl_cat}).
  Given a type family $V$ and $t\in T$ we set $V_t := V(t)$.
  A morphism $f : V \to W$ between two type families $V$ and $W$ is a family of maps indexed by $T$, 
  \[ f : t \mapsto f_t := f(t) : V_t \to W_t \enspace . \]
\end{definition}

\begin{remark}
 Equivalently to Def.\ \ref{def:category}, a category $\C$ is given by
  \begin{itemize}
   \item a collection $\C_0$ of objects and a collection $\C_1$ of morphisms,
   \item two maps 
       \[ \src, \tgt : \C_1 \to \C_0 \]
   \item a partially defined composition function 
         \[ (\comp{\_}{\_}) : \C_1 \times \C_1 \to \C_1 \enspace , \]
     such that $\comp{f}{g}$ is defined only for \emph{composable morphisms} $f$ and $g$, 
     i.e.\ if $\tgt(f) = \src(g)$. In this case we require that
            $\src(\comp{f}{g}) = \src(f)$ and $\tgt(\comp{f}{g}) = \tgt(g)$,
   \item identity morphisms and properties analoguous to those of the preceding definition. 
       The associative law, e.g., reads as
       \[ \forall f~g~h : \C_1, \enspace \tgt(f) = \src(g) \Longrightarrow \tgt(g) = \src(h) \Longrightarrow
                          \comp{f}{(\comp{g}{h})} = \comp{(\comp{f}{g})}{h}
       \]
  \end{itemize}
\end{remark}

\subsubsection{Which Definition to Formalize -- Dependent Hom--Types?}

The main difference w.r.t.\ formalization between these two definitions is that
of \emph{composability of morphisms}.
The first definition can be implemented directly only in type theories featuring \emph{dependent types},
such as the Calculus of Inductive Constructions (CIC). The ambient type system, i.e.\ the prover,
then takes care of composability -- terms with compositions of non--composable morphisms are 
rejected as ill--typed terms.

The second definition can be implemented also in provers with a simpler type system such as the 
family of HOL theorem provers. However, since those (as well as the CIC) are theories where functions are total,
one is left with the question of how to implement composition.
Composition might then be implemented either as a functional relation or as a total function
about which nothing is known (deducible) on non--composable morphisms.
The second possibility is implemented in O'Keefe's development \cite{OKeefe}.
There the author also gives an overview over available formalizations
in different theorem provers with particular attention to the choice of the 
definition of category.

In our favourite prover \textsf{Coq}, both definitions have been employed in significant
developments: the second definition is used in Simpson's construction of the Gabriel--Zisman
localization \cite{Simpson_localization}, whereas Huet and Sa\"ibi's \textsf{ConCaT} \cite{concat}
uses type families of morphisms as in the first definition.
To our knowledge there is no library in a prover with dependent types such as \textsf{Coq} or \textsc{NuPrl} \cite{nuprl}
 which develops and compares
both definitions w.r.t.\ provability, readability etc.

We decided to construct our library using type families of morphisms. 
In this way the proof of composability of two morphisms is done by \textsf{Coq}
type computation automatically.

\textsf{Coq}'s \emph{implicit argument} mechanism allows us 
to omit the deducible arguments, as we do in Def.\ \ref{def:category}
for the ``object arguments'' $c,d$ and $e$ of the composition.
Together with the possibility to define infix notations 
this brings our formal syntax close to informal mathematical syntax.

\subsubsection{Setoidal Equality on Morphisms}

All the properties of a category $\C$ concern equality of two parallel morphisms, i.e. morphisms with same source and target.
In \textsf{Coq} there is a polymorphic equality, called \emph{Leibniz equality}, readily available for any type.
However, this equality actually denotes \emph{syntactic equality}, which already in the case of maps 
does not coincide with the ``mathematical'' equality on maps -- given by pointwise equality -- that we would rather consider.
With the use of axioms -- for the mentioned example of maps the axiom \lstinline!functional_extensionality! from 
the \textsf{Coq} standard library -- 
one can often deduce Leibniz equality from the ``mathematical equality'' in question.
But this easily gets cumbersome, in particular when the morphisms -- as will be in our case -- 
are sophisticated algebraic structures composed of a lot of data and properties.
Instead, we require any collection of morphisms $\C(c,d)$ for objects $c$ and $d$ of $\C$
to be equipped with an equivalence relation, which plays the r\^ole of 
equality on this collection. In the \textsf{Coq} standard library 
equivalence relations are implemented as a type class with the 
underlying type as a parameter \lstinline!A!, 
and the relation as well as a proof of it being an equivalence as fields:
\begin{lstlisting}[caption={Setoid type class}, label={lst:setoid}]
Class Setoid A := {
  equiv : relation A ;
  setoid_equiv :> Equivalence equiv }.
\end{lstlisting}

Setoids as morphisms of a category have been used by Aczel \cite{aczel_galois} in LEGO (there a setoid is simply called ``set'')
and Huet and Sa\"ibi (HS) \cite{concat} in \textsf{Coq}.
HS's setoids are implemented as records of which the underlying type is a component
instead of a parameter. This choice makes it necessary to duplicate 
the definitions of setoids and categories in order to make them available with a ``higher'' type
\footnote{In HS's \textsc{ConCaT}, a type \lstinline!T! which is defined after the type of setoids cannot be the carrier of a setoid
 itself. What is done in HS's library is to define a type \lstinline!Setoid'! isomorphic to \lstinline!Setoid!
  \emph{after} the definition of \lstinline!T!. The type of \lstinline!Setoid'! now being higher 
  than that of \lstinline!T!, one can define an element of this type whose carrier is \lstinline!T!.}.

\subsubsection{\textsf{Coq} Setoids and their morphisms}\label{sec:setoids}

Setoids in \textsf{Coq} are implemented as a type class (cf.\ Listing \ref{lst:setoid})
with a type parameter \lstinline!A!
and a relation on \lstinline!A! as well as a proof of this relation being an equivalence
as fields. For the term \lstinline!equiv a b! the infix notation ``\lstinline!a == b!''
is introduced. The instance argument of \lstinline!equiv! is implicit (cf.\ Sec.\ \ref{sec:alg_structure}).

A \emph{morphism of setoids} between setoids \lstinline!A! and 
\lstinline!B! is a \textsf{Coq} function, say \lstinline!f!, on the underlying types
which is compatible with the setoid relations on the source and target. That is, 
it maps equivalent terms of \lstinline!A! to equivalent terms of \lstinline!B!, or, in mathematical notation,
\begin{equation} a \equiv_A a' \enspace \Longrightarrow \enspace f(a) \equiv_B f(a') \enspace . \label{eq:proper} \end{equation}
In the \textsf{Coq} standard library such morphisms are implemented as a type class
\begin{lstlisting}
Class Proper {A} (R : relation A) (m : A) : Prop :=
  proper_prf : R m m.
\end{lstlisting}
where the type \lstinline!A! is instantiated with a function type \lstinline!A -> B! and the 
relation \lstinline!R! on \lstinline!A -> B! is instantiated with pointwise compatibility
\footnote{
In the \textsf{Coq} standard library the definition of \lstinline!respectful! is actually a special case of 
a more general definition of a heterogeneous relation \lstinline!respectful_hetero!.
}
:
\begin{lstlisting}
Definition respectful {A B : Type} (R : relation A) (R' : relation B) : 
           relation (A -> B) :=
  fun f g => forall x y, R x y -> R' (f x) (g y).
Notation " R ==> R' " := (@respectful _ _ (R%signature) (R'%signature))
    (right associativity, at level 55) : signature_scope.
\end{lstlisting}
%
Given \textsf{Coq} types \lstinline!A! and \lstinline!B! equipped with relations 
\lstinline!R : relation A! and \lstinline!R' : relation B!, resp., and a map
\lstinline!f : A -> B!, 
the statement \lstinline!Proper (R ==> R') f! --- replacing aforementioned notation --- really means
\begin{lstlisting}
Proper (respectful R R') f , 
\end{lstlisting}
which is the same as \lstinline!respectful R R' f f!,
which itself just means
\begin{lstlisting}
forall x y, R x y -> R' (f x) (f y) .
\end{lstlisting}
This is indeed the statement of Display \eqref{eq:proper} in the special case that \lstinline!R!
and \lstinline!R'! are \emph{equivalence} relations.

For any component of an algebraic structure that is a map defined on setoids,
we add a condition of the form \lstinline!Proper...! in the formalization.
Examples are the categorical composition (Lst.\ \ref{lst:cat}) and the 
monadic substitution map (Lst.\ \ref{lst:monad}).
Rewriting related terms under those equivalence relations is tightly integrated in the \lstinline!rewrite!
tactic of \textsf{Coq}.

\subsubsection{\textsf{Coq} implementation of categories}\label{sec:coq_impl_cat}

Finally we adopt Sozeau's definition of category \cite{sozeau.Coq/classes/fctc}, 
which itself is a type class version of the definition given by Huet and Sa\"ibi \cite{concat}.
The type class of categories is parametrized by a type of objects and a type family of morphisms, whose parameters are the source and target objects.
\begin{lstlisting}[label={lst:cat}, caption={Type class of categories}]
Class Cat_struct (obj : Type)(mor : obj -> obj -> Type) := {
  mor_oid :> forall a b, Setoid (mor a b) ;
  id : forall a, mor a a ;
  comp : forall {a b c}, mor a b -> mor b c -> mor a c ;
  comp_oid :> forall a b c, Proper (equiv ==> equiv ==> equiv) (@comp a b c) ;
  id_r : forall a b (f: mor a b), comp f (id b) == f ;
  id_l : forall a b (f: mor a b), comp (id a) f == f ;
  assoc : forall a b c d (f: mor a b) (g:mor b c) (h: mor c d),
      comp (comp f g) h == comp f (comp g h) }.
\end{lstlisting}
Compared to the informal definition \ref{def:category} there are two additional fields:
the field \lstinline!mor_oid! of type \lstinline!forall a b, Setoid (mor a b)! equips each collection of morphisms \lstinline!mor a b! with
a custom equivalence relation.
The field \lstinline!comp_oid! states that the composition \lstinline!comp! of the category is compatible with the setoidal structure on the morphisms
given by the field \lstinline!mor_oid! as explained in Sec.\ \ref{sec:setoids}.
We recall that setoidal equality is overloaded and denoted by the infix symbol `\lstinline!==!'. 
In the following we write `\lstinline!a ---> b!' for \lstinline!mor a b! and \lstinline!f;;g! 
for the composition of morphisms \lstinline!f : a ---> b! and \lstinline!g : b ---> c!
\footnote{\textsf{Coq} deduces and inserts the missing ``object'' arguments \lstinline!a!, \lstinline!b! and \lstinline!c! of 
the composition automatically from the type of the morphisms. 
For this reason those object arguments are called \emph{implicit} (cf.\ Sec.\ \ref{sec:alg_structure}).}.

The implementation of the category $\TS{T}$ of Def.\ \ref{def:TST} uses \textsf{Coq} types as sets:
(the properties being proved automatically by a suitable tactic invoked by the \lstinline!Program! framework, cf.\ Subsec.\ \ref{subsec:program}):
\begin{lstlisting}
Program Instance ITYPE_struct : Cat_struct (obj := T -> Type)
          (fun A B => forall t, A t -> B t) := {
  mor_oid := INDEXED_TYPE_oid ;  (* pointwise equality in each component of the family of maps *)
  comp A B C f g := fun t => fun x => g t (f t x) ;
  id A := fun t x => x }.
\end{lstlisting}
The objects of this category are hence implemented as families of \textsf{Coq} types, indexed by a fixed \textsf{Coq} type \lstinline!T!.
Morphisms between two such objects are suitable families of \textsf{Coq} functions.


\subsubsection{\texorpdfstring{Interlude on the \lstinline!Program! feature}{Interlude on the Program feature}}\label{subsec:program}

The \lstinline!Program Instance! vernacular allows to fill in fields of an instance of a type class by means of tactics.
Indeed, when omitting a field in an instance declaration --- such as the proofs of associativity \lstinline!assoc! 
and left and right identity \lstinline!id_l! and \lstinline!id_r! in 
the instance \lstinline!ITYPE_struct! in the previous listing --- the \lstinline!Program! framework creates an \emph{obligation}
for each missing field, making use of the information that the user provided for the other fields.
As an example, the obligation created for the field \lstinline!assoc! of the previous example is to prove associativity for the composition defined by
\begin{lstlisting}
comp f g := fun t => fun x => g t (f t x) .
\end{lstlisting}
It then tries to solve the resulting obligations using the tactic that the user has specified via the \lstinline!Obligation Tactic!
command. In case the automatic resolution of the obligation fails, the user can enter the interactive proof mode finish the proof manually.

It is technically possible to fill in both \emph{data} and \emph{proof} fields automatically via the \lstinline!Program! framework. 
However, in order to avoid the automatic inference of data which we cannot control,
we always specify \emph{data} directly as is done in the case of \lstinline!ITYPE_struct!, and rely on automation via \lstinline!Program! only for proofs.


\subsection{Invertible morphisms, Initial objects}

Given a category $\C$, a morphism $f : c \to d$ from object $c$ to object $d$ is called \emph{invertible},
if there exists a left-- and right--inverse $g : d \to c$, that is, a morphism $g : d \to c$
such that $\comp{f}{g} = id_c$ and $\comp{g}{f} = id_d$.
In this case the objects $c$ and $d$ are called \emph{isomorphic}.

An initial object of a category is an object for which there is precisely \emph{one} morphism to
any object of the category:
\begin{definition} \label{def:initial_object}
  Let $\C$ be a category, and $c \in \C$ an object of $\C$. 
    The object $c$ is called \emph{initial} if for any object $d \in \C$ there exists 
  a unique morphism $i_d : c\to d$ from $c$ to $d$ in $\C$.
\end{definition}

\begin{remark}\label{rem:initial_unique}
It is easy to see that any two initial objects of a category $\C$ are isomorphic via a unique isomorphism. This justifies the use 
of the definite article, i.e.\  speaking about ``\emph{the}'' initial object of a category --- if it exists.
\end{remark}
Formally, we implement the initiality structure as a type class which inherits from the class of categories.
Its fields are given by an object \lstinline!Init! of the category, a map \lstinline!InitMor! mapping
each object \lstinline!a! of the category to a morphism from \lstinline!Init! to \lstinline!a! 
and a proposition stating that \lstinline!InitMor a! is unique for any object \lstinline!a!.
\begin{lstlisting}
Variable ob : Type.
Variable mor : ob -> ob -> Type.
Class Initial (C : Cat_struct mor) := {
  Init : ob;
  InitMor: forall a : ob, mor Init a;
  InitMorUnique: forall a (f : mor Init a), f == InitMor a }.
\end{lstlisting}
Note that the initial morphism is \emph{not} given by an existential statement of the form $\forall a, \exists f : \ldots$,
or, in \textsf{Coq} terms, using an \lstinline!exists! statement.
This is because the \textsf{Coq} existential lies in \lstinline!Prop! and hence does not allow
for elimination -- witness extraction -- when building anything but proofs.

\subsection{Functors \& Natural Transformations}

Given two categories $\C$ and $\D$, a functor $F:\C\to\D$ maps objects of $\C$ to objects of $\D$, and 
morphisms of $\C$ to morphisms of $\D$, while preserving source and target:

\begin{definition} A functor $F$ from $\C$ to $\D$ is given by
 \begin{itemize}
  \item a map $F : \C \to \D$ on the objects of the categories involved and
  \item for any pair of objects $(c,d)$ of $\C$, a map
     \[ F_{(c,d)} : \C(c,d) \to \D(Fc, Fd) \enspace , \]
 \end{itemize}
 such that
  \begin{itemize}
   \item $\forall c : C, \enspace F(id_c) = id_{Fc}$ and 
   \item $\forall c~d~e : C, \forall f:c\to d, \forall g:d\to e, \enspace F(\comp{f}{g}) = \comp{Ff}{Fg}$.
  \end{itemize}
Here we use the same notation for the map on objects and that on morphisms. For the latter we also omit the 
subscript ``$(c,d)$'' as instances of implicit arguments. For its implementation we refer to the \textsf{Coq} source files.
\end{definition}

\begin{definition}
  Let $F,G : \C \to \D$ be two functors from $\C$ to $\D$. A \emph{natural transformation} $\tau : F \to G$ 
  associates to any object $c \in \C$ a morphism
    \[  \tau_c : Fc \to Gc \]
  such that for any morphism $f : c \to d$ in $\C$ the following diagram commutes:
 \[
  \begin{xy}
   \xymatrix{   Fc \ar[r]^{\tau_c} \ar[d]_{Ff} & Gc \ar[d]^{Gf} \\
                Fd \ar[r]_{\tau_d} & Gd
}
  \end{xy} 
 \]

\end{definition}

\subsection{Monads, modules and their morphisms}

Monads have long been known to capture the notion of substitution, cf.\ \cite{alt_reus}. 
The closely connected notion of \emph{module over a monad} was recently introduced in the context of abstract syntax 
by Hirschowitz and Maggesi \cite{DBLP:conf/wollic/HirschowitzM07}.
Similarly to the two equivalent definitions of monads as presented by Manes \cite{manes} there are two equivalent definitions of modules over a monad. 
Contrary to the given reference \cite{DBLP:conf/wollic/HirschowitzM07}  we use the definition of monad as a \emph{Kleisli triple},
since this definition is well--known for its use in the functional programming language \textsc{Haskell} 
and hence accessible to a relatively wide audience.

\begin{definition}\label{def:monad} A \emph{monad} $P$ over a category $\C$ is given by 
\begin{itemize}
 \item 
a map $P\colon \C \to\C$ on the objects of $\C$ (by abuse of notation it carries the same name as the monad),
 \item 
for each object $c$ of $\C$, a morphism $\eta_c\in \C(c,Pc)$ and
 \item 
for all objects $c$ and $d$ of $\C$ a \emph{substitution} map 
 \[\sigma_{c,d}\colon \C (c,Pd) \to \C (Pc,Pd)\]
\end{itemize}
 such that the following diagrams commute for all suitable morphisms $f$ and $g$:
\begin{equation*}
\begin{xy}
\xymatrix @=3pc{
c \ar [r] ^ {\we_c} \ar[rd]_{f} & Pc \ar[d]^{\kl{f}} & Pc \ar@/^1pc/[rd]^{\kl{\we_c}} \ar@/_1pc/[rd]_{\id}& {} &
     Pc \ar[r]^{\kl{f}} \ar[rd]_{\kl{\comp{f}{\kl{g}}}} & Pd\ar[d]^{\kl{g}}\\
{} & Pd , & {} & Pc , &{} & Pe .\\
}
\end{xy}
\end{equation*}
  We omit the subscripts of the substitution map as done in the diagrams.
\end{definition}

\begin{example}[Lists]
  Consider the map $[\_] : \Set \to \Set$ mapping any set $X$ to the set $\listop(X)$ of lists over $X$, together 
  with the following maps:
  \begin{lstlisting}
Definition eta (X : Type) (x : X) := x::nil.    (* the singleton list *)

Fixpoint sigma X Y (f : X -> list Y) (l : list X) :=
     match l with nil => nil | x::l' => app (f x) (sigma f l') end.    (* app = append *)
  \end{lstlisting}
This defines a monad structure on lists, the axioms are easily verified.
%
\end{example}



\begin{example} \label{ex:ring_monad}
 Let $\mathbf{R}$ be a commutative ring. To any set $X$ we associate the set $R(X)$ of polynomials with variables in $X$ and coefficients in $\mathbf{R}$:
 \[   R : X \mapsto R(X) \enspace . \]
 We equip the map $R$ with a monad structure by defining the unit $\eta$ as
 \[ \eta_X : x \mapsto x \enspace \text{ (considered as a polynomial)} \enspace . \]
 The monad substitution is best defined using two auxiliary functions: 

firstly, for $f : X \to Y$, we set 
\[ R(f) : R(X) \to R(Y)\enspace , \quad p(x_1,\ldots,x_n) \mapsto p(f(x_1), \ldots, f(x_n)) \enspace , \]
yielding a functor with object map $X \mapsto R(X)$.

Secondly, for any set $X$, we define a \emph{multiplication} 
\[ \mu_X : R(R(X)) \to R(X) \]
which, given a polynomial $p\bigl(p_1(x_1,\ldots,x_n), \ldots,p_m(x_1,\ldots,x_n)\bigr)$ with polynomials as variables, allows
to consider it as a polynomial $p(x_1,\ldots,x_n)$ after expansion. Here we can suppose all polynomials $p_i$ to have variables 
 in the same finite set $\{x_1,\ldots,x_n\}$.
 The substitution map is then defined using those auxiliary maps:
 \begin{equation} \label{eq:kl_from_mu}
\sigma_{X,Y} : (X \to R(Y)) \to R(X) \to R(Y)\enspace , \quad \sigma_{X,Y}(f)(x) := \comp{R(f)}{\mu_Y} \enspace .   
 \end{equation}
Later (cf.\ Def.\ \ref{def:module}) we define the notion of \emph{module} over a monad. 
In Ex.\ \ref{ex:module_ring} we show how any module over $\mathbf{R}$ in the classical sense gives rise to a module over $R$ in the sense of
Def.\ \ref{def:module}.
\end{example}

\begin{remark}
 The preceding example actually illustrates a use of the aforementioned equivalent definition of monad as a triple $(T,\eta,\mu)$
 where $T$ is an endofunctor on a category $\C$ and $\eta : \Id \to T$ and $\mu : TT\to T$ are natural transformations verifying
 some properties. Display \eqref{eq:kl_from_mu} indicates how to define the monad substitution $\sigma$ from monad multiplication $\mu$.
 We refer to \cite{manes} for details.
\end{remark}

\begin{remark}
 Let $A$ be an algebra over the ring $\mathbf{R}$ of Ex.\ \ref{ex:ring_monad}. 
 Then $A$ is an $R$--algebra (we refer to \cite{maclane} for the definition): 
the map $\alpha:R(A) \to A$ is induced by the module operation $\phi : R\times A \to A$ and 
the bilinear product on $A$.
   The commutation properties of the following diagrams is a consequence of the rules the module operation $\phi$ verifies.
 \[
  \begin{xy}
   \xymatrix@C=3pc{ R(R(A))  \ar[d]_{\mu^R_A} \ar[r]^{R\alpha} & R(A) \ar[d]^{\alpha} \\
               R(A) \ar[r]_{\alpha} & A
}
  \end{xy}\qquad
  \begin{xy}
   \xymatrix{   A \ar[dr]_{\id} \ar[r]^{\eta_A }& R(A)\ar[d]^{\alpha} \\
                {} & A
}
  \end{xy}
 \]

\end{remark}



\begin{example}\label{ex:ulc_monad}(Ex.\ \ref{ex:ulc_def} cont.)
  This example is due to Altenkirch and Reus \cite{alt_reus}.
  We consider the map $\LC$ associating to any set $X$ the set of 
  untyped lambda terms with free variables in $X$. 
  Given any set $X$, the constructor $\Var(X) : X \to \LC(X)$
  maps a variable to itself, this time seen as a lambda term.
 The substitution map is defined recursively, 
  using a helper function \lstinline!shift! 
  when going under the binding constructor $\Abs$:
\begin{lstlisting}
Fixpoint subst V W (f : V -> LC W) (y : LC V) : LC W :=
  match y in LC _ return LC W with
  | Var v => f v
  | Abs v => Abs (subst (shift f) v)
  | App s t => App (subst f s) (subst f t)
  end.
\end{lstlisting}
The function \lstinline!shift! is of type $\shiftop_{V,W}:(V \to \LC(W)) \to V^* \to \LC(W^*)$, sending the additional
variable of $V^*$ to $\Var(*_W)$.
These definitions yield a monad $\LC$ with $\eta := \Var$ and $\mu := $ \lstinline!subst!.
\end{example}

\begin{example}\label{ex:slc_monad}
  Consider the simply--typed lambda calculus as in Ex.\ \ref{ex:slc_def}.
  Definitions similar to those of Ex.\ \ref{ex:ulc_monad}, but additionally 
  indexed by object types of $\mathcal{T}$, turn $\SLC$ into a monad on the category
  $\TS{\mathcal{T}}$. The definition of the substitution map $\sigma$ reads as follows:
\begin{lstlisting}
Fixpoint subst (V W : IT) (f : V ---> TLC W) t (y : TLC V t) : TLC W t :=
  match y with
  | Var _ v => f _ v
  | Abs _ _ v => Abs (subst (shift f) v)
  | App _ _ u v => App (subst f u) (subst f v)
  end.
\end{lstlisting}
where the object type arguments are partially implicit and otherwise denoted by the underscore ``\lstinline!_!'' in 
the pattern matching branches. 
The \lstinline!shift! map is -- similarly to the preceding, untyped example -- necessary to adapt the substitution map $f$
to the enlarged domain and codomain under binders (cf.\ Sec.\ \ref{sec:deriv}).
\end{example}

\begin{example}\label{ex:derived_monad}
  For any set $X$, let $X^* := X \amalg \{*\}$.
 Given any monad $P$ on the category of sets, the map $P^* : X \mapsto P(X^*)$ inherits a monad structure from $P$.
In detail, a monadic substitution for $P^*$ is defined, for a morphism $f : X \to P^*(Y)$, as
 \[ \kl[P^*]{f} := \kl[P]{\defaultmap(f,\eta_{Y^*}(*))} \enspace .  \]
The map \[\defaultmap(f, \eta_{Y^*}(*)) : X^* \to P^*(Y)\] sends the additional variable $*$ to $\eta(*)$.
\end{example}


\noindent
Given a monad $P$ over $\C$ and a morphism $f:c \to d$ in $\C$, we define %
\[
   P(f) := \lift_P(f) := \kl{\comp{f}{\we_d}}\enspace ,
\]
thus equipping $P$ with a functorial structure (\lstinline!lift!). In case $P$ is a syntax, e.g.\ the monad $\LC$ of
Ex.\ \ref{ex:ulc_monad}, the \lstinline!lift! operation corresponds to \emph{variable renaming} according to the map $f$.
Note that $f$ is not necessarily bijective, and hence $P(f)$ not necessarily a \emph{permutation} of variables.

The formal definition of monad is almost a literal translation of Def.\ \ref{def:monad}. 
The only difference is an additional field \lstinline!kleisli_oid! stating that
the substitution map is a map of setoids (cf.\ Sec.\ \ref{sec:setoids}):
\begin{lstlisting}[caption={Type class of monads}, label={lst:monad}]
Class Monad_struct (C : Cat) (F : C -> C) := {
  weta : forall c, c ---> F c ;
  kleisli : forall a b, (a ---> F b) -> (F a ---> F b) ;
  kleisli_oid :> forall a b, Proper (equiv ==> equiv) (kleisli (a:=a) (b:=b)) ;
  eta_kl : forall a b (f : a ---> F b), weta a ;; kleisli f == f ;
  kl_eta : forall a, kleisli (weta a) == id _ ;
  dist : forall a b c (f : a ---> F b) (g : b ---> F c),
      kleisli f ;; kleisli g == kleisli (f ;; kleisli g) }.
\end{lstlisting}
As in the informal Def.\ \ref{def:monad} the ``object'' arguments of the substitution map \lstinline!kleisli! 
are implicit.

For two monads $P$ and $Q$ over the same category $\C$ a \emph{morphism of monads} is a 
family of morphisms $\tau_c\in\C(Pc,Qc)$ 
that is compatible with the monadic structure:

\begin{definition}\label{def_monad_hom}
 A \emph{morphism of monads} (\lstinline!Monad_Hom!) from $P$ to $Q$ is given by a collection of morphisms $\tau_c\in \C(Pc,Qc)$ 
such that the following diagrams commute for any morphism $f : c \to Pd$:
\begin{equation*}
 \begin{xy}
  \xymatrix @=3pc{
  Pc \ar[r]^{\kl[P]{f}} \ar[d]_{\tau_c}& Pd \ar[d]^{\tau_d} & c \ar[r]^{\we^P_c} \ar[rd]_{\we^Q_c} & Pc \ar[d]^{\tau_c} \\
  Qc \ar[r]_{\kl[Q]{\comp{f}{\tau_d}}} & Qd  , & {} & Qc .
}
 \end{xy}
\end{equation*}
\end{definition}
Two monad morphisms are said to be equal if they are equal on each object.

The formal definition is a straightforward transcription, even if the diagrams do not read as nicely there: 
\begin{lstlisting}
Class Monad_Hom_struct (Tau: forall c, P c ---> Q c) := {
  monad_hom_kl: forall c d (f: c ---> P d),
      kleisli f ;; Tau d == Tau c ;; kleisli (f ;; Tau d) ;
  monad_hom_weta: forall c: C, weta c ;; Tau c == weta c }.
\end{lstlisting}
Observe that some arguments are inferred by \textsf{Coq}, such as to which monad the respective \lstinline!kleisli! and \lstinline!weta! operations belong.

It follows from these commutativity properties that the family $\tau$ is a natural transformation between the functors induced by the monads $P$ and $Q$. 
Monads over $\C$ and their morphisms form a category \lstinline!MONAD C! where identity and composition of morphisms are 
simply defined by pointwise identity resp.\ composition of morphisms:
\begin{lstlisting}[label={lst:monad_comp_id}, caption={Composition and identity for monad morphisms}]
Variables P Q R : Monad C.
Variable S : Monad_Hom P Q.
Variable T : Monad_Hom Q R.
Instance Monad_Hom_comp_struct : Monad_Hom_struct (fun c => S c ;; T c).
Instance Monad_Hom_id_struct : Monad_Hom_struct (fun c => id (P c)).
\end{lstlisting}


We illustrate the concept of monad morphism by showing how abstraction fails to be such 
a morphism.
The map $V \mapsto \LC(V)$ is object function of a monad, as is the map $\LC^*:V \mapsto \LC(V^*)$ (cf.\ Ex.\ \ref{ex:derived_monad}).
However, the constructor $\Abs$, while having the suitable type, is not a morphism of monads from $\LC^*$ to $\LC$;
it does not verify the square diagram of Def.\ \ref{def_monad_hom}:
\begin{example}\label{ex:abs_not_monadic}
   The following diagram \emph{fails} to commute for the map 
     \[f : a \mapsto \Var (*) \enspace ; \]
the term $\Var(a)\in \LC(\{a\})$ maps to $\lambda x. x$
when taking the upper route, while mapping to $\lambda xy.y$ when taking the lower route:
\begin{equation} \label{eq:failing_abs}
 \begin{xy}
 \xymatrix{    **[l]  \Var(a)\in\LC^*(\{a\}) \ar[dd]_{\Abs_{\{a\}}} \ar[r]^{\kl[\LC^*]{f}} & **[r] \LC^* (\emptyset) \ar[d]^{\Abs_Y} \\
                          {}      & **[r] \LC^*(\emptyset) \ni  \lambda x . x              \\
               **[l]  \LC(\{a\})  \ar[r]_{\kl{\comp{f}{\Abs_Y}}}   &               **[r] \LC(\emptyset) \ni \lambda xy.y
}
 \end{xy}
\end{equation}
This is due to the additional abstraction appearing through the
   lower vertical substitution morphism.

Instead, we will equip the constructor $\Abs$ with the structure of a \emph{module morphism} (Def.\ \ref{def:mod_hom}), 
cf.\ Exs.\ \ref{ex:ulc_taut_mod}, \ref{ex:ulc_const_mod} and \ref{ex:ulc_mod_mor}.
Module morphisms verify a diagram similar to the square diagram of monad morphisms, with the difference that the underlying 
natural transformation (here $\Abs$) does not appear in the lower vertical substitution.

\end{example}

The preceding example for the constructor $\Abs$ shows the need for a concept
that is more general than that of monads and monad morphisms,
while still expressing compatibility of the underlying natural transformation with substitution.

For this reason, we consider \emph{modules over monads}, which generalize the notion of monadic substitution, and \emph{module morphisms}:


\begin{definition}\label{def:module}
Let $\D$ be a category. A \emph{module $M$ over $P$ with codomain $\D$} is given by
\begin{itemize}
 \item a map $M\colon \C \to \D$ on the objects of the categories involved and 
 \item for all objects $c,d$ of $\C$ a map 
      \[ 
          \varsigma_{c,d}\colon \C (c,Pd) \to \C (Mc,Md)
      \]
\end{itemize}
such that the following diagrams commute for all suitable morphisms $f$ and $g$:
\begin{equation*}
\begin{xy}
\xymatrix @=3pc{
 Mc \ar[r]^{\mkl{f}} \ar[rd]_{\mkl{\comp{f}{\kl{g}}}} & Md\ar[d]^{\mkl{g}} & Mc  \ar@/^1pc/[rd]^{\mkl{\we_c}} \ar@/_1pc/[rd]_{\id} & {} \\
  {} & Me ,& {} & Mc .\\
}
\end{xy}
\end{equation*}
\end{definition}
\noindent
A functoriality for such a module $M$ is then defined similarly to that for monads (\lstinline!mlift!):
\[M(f) :=  \mlift_M(f) := \mkl{\comp{f}{\we^P}}\enspace .\]



\begin{example}(Ex.\ \ref{ex:ring_monad} cont.) \label{ex:module_ring}
Let $\mathbf{R}$ be a commutative ring. For any set $X$, $R(X)$ is a module over $\mathbf{R}$ in the classical, algebraic sense.
Let $\mathbf{M}$ be any module over $\mathbf{R}$.
We define a map 
\[ M : X \mapsto M(X) := \mathbf{M} \otimes_{\mathbf{R}} R(X) \enspace , \]
 where $\_\otimes_{\mathbf{R}}\_$ denotes the tensor product of modules. We omit the index $\mathbf{R}$ of the tensor product.
This map is the object function of a module (in the sense of Def.\ \ref{def:module}) 
over the monad $R$ (cf.\ Ex.\ \ref{ex:ring_monad}).
The module substitution is defined using the fact that the tensor product is functorial in the second argument:
\[ \varsigma_{X,Y} : (X \to R(Y)) \to \mathbf{M}\otimes R(X) \to \mathbf{M}\otimes R(Y) \enspace , \quad f \mapsto \mathbf{M}\otimes \sigma_{X,Y}(f) \enspace . \]
\end{example}


The implementation of modules resembles that of monads:

\begin{lstlisting}
Class Module_struct (M : C -> D) := {
  mkleisli: forall c d, (c ---> P d) -> (M c ---> M d);
  mkleisli_oid :> forall c d, 
      Proper (equiv ==> equiv) (mkleisli (c:=c)(d:=d));
  mkl_weta: forall c, mkleisli (weta c) == id _ ;
  mkl_mkl: forall c d e (f : c ---> P d) (g : d ---> P e),
      mkleisli f ;; mkleisli g == mkleisli (f ;; kleisli g) }.
\end{lstlisting}

We anticipate several constructions on modules to give some further examples of modules:

\begin{example}\label{ex:ulc_taut_mod}(Ex.\ \ref{ex:ulc_monad} cont.)
   Any monad $P$ on a category $\C$ can be considered as a module over itself, the \emph{tautological module} (cf.\ Sec.\ \ref{mod_examples}).
   In particular, the untyped lambda calculus $\LC$ is a $\LC$--module with codomain $\Set$.
\end{example}

\begin{example}\label{ex:ulc_const_mod}
  The map  \[ \LC^* : V \mapsto \LC(V^*) \]
  can be equipped with a structure as $\LC$--module, the \emph{derived module} of (the module) $\LC$ (cf.\ Sec.\ \ref{sec:deriv}).
  Also, the map \[\LC\times \LC : V \mapsto \LC(V)\times\LC(V)\] can be equipped with a structure as $\LC$--module.
\end{example}

\begin{example}\label{ex:slc_modules}
   Consider the monad $\SLC : \TS{\mathcal{T}} \to \TS{\mathcal{T}}$ of Ex.\ \ref{ex:slc_monad}.
Given any object type $t\in \mathcal{T}$, the map 
  \begin{equation}\SLC_t : V \mapsto \SLC(V)_t \label{eq:slc_fib}\end{equation}
  can be equipped with the structure of a module over $\SLC$ with codomain category $\Set$ (cf.\ Sec.\ \ref{sec:fibre}).
 Similarly, for $s\in \mathcal{T}$, the map \[\SLC^{s} : V \mapsto \SLC(V^{*s})\]
 can be equipped with a module structure over the monad $\SLC$ (cf.\ Sec.\ \ref{sec:deriv}). %

Those two operations, fibre and derivation, can be combined, yielding a module over $\SLC$ with carrier
\[ V\mapsto \SLC^{s}_t(V) := \SLC(V^{*s})_t \enspace .\]

The final example is that of products: the map
\[\SLC_{s\Rightarrow t} \times\SLC_{s} : V \mapsto \SLC(V)_{s\Rightarrow t} \times \SLC(V)_{s}\]
  can be equipped with the structure of a module (cf.\ Sec.\ \ref{mod_examples}).
\end{example}

Those three constructions are our main examples of modules. From the last example the reader may have guessed that
we will consider the domain and codomain of some constructor to be given as modules:
here the domain of (an uncurried version of) the constructor $\App_{s,t}$ (cf.\ Ex.\ \ref{ex:slc_def}) of the simply--typed
lambda calculus is a module over $\SLC$ with codomain $\Set$.
The constructors themselves then are \emph{morphisms of modules}:

\begin{definition}\label{def:mod_hom} %
 Let $M$ and $N$ be two modules over $P$ with codomain $\D$. A \emph{morphism of $P$--modules} from $M$ to $N$ is given by a collection of morphisms $\rho_c\in\D(Mc,Nc)$ such that for 
any morphism $f\in \C(c,Pd)$ the following diagram commutes:
\begin{equation*}
 \begin{xy}
  \xymatrix @=3pc{
  Mc \ar[r]^{\mkl[M]{f}} \ar[d]_{\rho_c}& Md \ar[d]^{\rho_d}  \\
  Nc \ar[r]_{\mkl[N]{f}} & Nd. \\
}
 \end{xy}
\end{equation*}
\end{definition}
We omit the formal definition.
A module morphism $M \to N$ also constitutes a natural transformation between the functors $M$ and $N$ induced by the modules.

\begin{example}\label{ex:ulc_mod_mor}(Ex.\ \ref{ex:ulc_const_mod} cont.)
  The map \[V \mapsto \App_V : \LC(V)\times \LC(V) \to \LC(V)\]
    verifies the diagram of the preceding definition and is hence a morphism of $\LC$--modules from $\LC\times\LC$ to $\LC$.
  The map \[ V \mapsto \Abs_V : \LC(V^*) \to \LC(V) \] is a morphism of $\LC$--modules from $\LC^*$ to $\LC$.
\end{example}

\begin{example}\label{ex:slc_mod_mor}(Ex.\ \ref{ex:slc_modules} cont.)
  Given $s,t\in \mathcal{T}$, the map 
    \[ \App(s,t) : V \mapsto 
          \App_V (s,t) : \SLC(V)_{s\Rightarrow t} \times \SLC(V)_{s} \to \SLC(V)_t 
    \]
  verifies the diagram of the preceding definition and is hence a morphism of modules 
 \[\SLC_{s\Rightarrow t} \times \SLC_{s} \to \SLC_t \enspace .\]

 In the same way the constructor $\Abs(s,t)$ is a morphism of modules from $\SLC^s_t$ to $\SLC_{s\Rightarrow t}$.
\end{example}

The modules over a monad $P$ and with codomain $\D$ and morphisms between them form a 
category called $\Mod^P_\D$ (in the library: \lstinline!MOD P D!), similar to the category of monads.

\subsection{Constructions on modules}\label{mod_examples}
The following constructions on monads and modules play a central role in what follows.

\begin{examp}{Tautological Module} {(\lstinline!Taut_Mod!)}
Every monad $P$ over $\C$ can be viewed as a module (also denoted by $P$) over itself, i.e.\ as an object in the category $\Mod^P_\C$:
\begin{lstlisting}
Program Instance Taut_Mod_struct : Module_struct P D P := {
  mkleisli c d f := kleisli (Monad_struct:=P) f;
  mkleisli_oid c d := kleisli_oid (a:=c)(b:=d);
  mkl_mkl c d e f g := dist f g;
  mkl_weta c := kl_eta (Monad_struct := P) c }.
\end{lstlisting}
In this definition we have actually inserted the section parameters \lstinline!P! and \lstinline!D!
of \lstinline!Module_struct! compared to the original code.
The second argument \lstinline!P! does not denote the monad $P$ but rather -- by coercion -- its underlying 
map on objects $P : \C\to \C$.
The fact that we call $P$ the monad as well as its tautological module is reflected formally in the coercion
\begin{lstlisting}
Coercion Taut_Mod : Monad >-> obj.
\end{lstlisting}
\end{examp}

\begin{examp}{Constant and terminal module} {(\lstinline!Const_Mod, MOD_Terminal!)} 
For any object $d \in \D$ the constant map $T_d\colon\C\to\D$, $c\mapsto d$ for all $c\in \C$ can be provided with the structure of a $P$--module for any monad $P$. 
In particular, if $\D$ has a terminal object $1_\D$, then the constant module $c \mapsto 1_\D$ is terminal in $\Mod^P_\D$.
\end{examp}

\begin{examp}{Pullback module} {(\lstinline!PbMod!)}
Given a morphism of monads $h : P \to Q$ and a $Q$-module $M$ with codomain $\D$, we define a $P$-module $h^* M$ with same object map $M: \C\to \D$ 
with substitution map \[\mkl[h^*M] f := \mkl[M]{\comp {f}{h_d}}.
        \] This module is called the \emph{pullback module of $M$ along $h$}. 
\begin{lstlisting}
Program Instance PbMod_struct (M : MOD Q D) : Module_struct P (D:=D) M := {
  mkleisli c d f := mkleisli (f ;; h d) }.
\end{lstlisting}
  The pullback extends to module morphisms (\lstinline!PbMod_Hom!) and is functorial.
\end{examp}


\begin{remark}\label{rem:about_pullback}
Note that pulling back the $Q$--module $M$ does not change the underlying functor. Similarly, pulling back a $Q$--module morphism $s : M \to M'$
does not modify the underlying natural transformation. It merely changes the substitution action: while the module substitution of 
$M$ takes morphisms $f : c \to Qd$ as arguments, the module $h^*M$ takes as arguments morphisms of the form $c \to Pd$.
\end{remark}

\begin{examp}{Induced module morphism} {(\lstinline!PbMod_ind_Hom!)}
With the same notation as in the previous example, the monad morphism $h$ induces a morphism of $P$--modules 
$h: P \to h^*Q$.
Again, in \textsf{Coq} we can indeed declare a 
\begin{lstlisting}
Coercion PbMod_ind_Hom : Monad_Hom >-> mor. 
\end{lstlisting} 
corresponding to above abuse of notation.
\end{examp}

\begin{remark}
 The module morphism $h$ induced by the monad morphism $h$ really consists of the same data, namely, for any object $c\in \C$,
  the morphism $h_c : Pc \to Qc$ in $\C$. In Sec.\ \ref{sec:mor_of_reps} we need to define the composite of a monad morphism 
  with a module morphism. This is done by considering, instead of the monad morphism, the module morphism it induces.
\end{remark}


\begin{examp}{Products} {(\lstinline!Prod_Mod!)}
Suppose the category $\D$ is equipped with a binary product. Let $M$ and $N$ be $P$--modules with codomain $\D$. We extend the map 
\[ \C\to\D, \quad c \mapsto Mc \times Nc \] 
to a module called the \emph{product of $M$ and $N$}:
\begin{lstlisting}
Program Instance Prod_Mod_struct : Module_struct (fun a => M a x N a) := {
   mkleisli c d f := (mkleisli f) X (mkleisli f) }.
\end{lstlisting}
This construction extends to a product on $\Mod^P_\D$.
For the implementation of binary product \lstinline!Cat_Prod! on a category, we refer to the library files.
\end{examp}

Our basic category of interest $[T,\Set]$ (in the library: \lstinline!ITYPE T!) is formalized as a category where objects are collections of {\sf Coq} types indexed by $T$.

The following two constructions -- fibre and derivation -- apply to monads and modules over the category of (families of) sets.

\subsection{Fibres} \label{sec:fibre}
For a module $M \in \Mod^P_{[T,\Set]}$ and $u\in T$, the \emph{fibre module} $M_u \in \Mod^P_{\Set}$ is defined by 
\[M_u V := (M V) (u) \] and \[ \mkl[M_u]{f} := \mkl[M]{f} (u) \enspace ,\]
that is, by forgetting all but one component of the indexed family of sets:
\begin{lstlisting}
Program Instance ITFibre_Mod_struct u : Module_struct P (fun c => M c u) := {
  mkleisli a b f := mkleisli (Module_struct := M) f u }.
\end{lstlisting}
The construction extends to a functor (\lstinline!ITFIB_MOD u!)
\[ (\_) _{u}\colon \Mod^P_{[T,\Set]} \to \Mod^P_{\Set} \enspace .\]

\subsection{Derivation}\label{sec:deriv}

Roughly speaking, a binding constructor makes free variables disappear. 
Its inputs are hence terms ``with (one or more) additional free variables'' compared to the output.

Let $T$ be a discrete category (a set) and $u\in T$ an element of $T$.
Define $D(u)$ to be the object of $\TS{T}$ such that 
\[D(u)(u)=\lbrace *\rbrace \quad\text{ and } \quad D(u)(t) = \emptyset \text{ for } t\neq u \enspace . \]
We enrich the object $V$ of $\TS{T}$ with respect to $u$ by setting
  \[ V^{*u} := V + D(u), \]
i.e.\ we add a fresh variable of type $u$.
Formally, we use an inductive type to construct this coproduct, in order to use pattern matching to define coproduct maps.
\begin{lstlisting}
Inductive opt (u : T) (V : ITYPE T) : ITYPE T :=
  | some : forall t : T, V t -> opt u V t
  | none : opt u V u.
\end{lstlisting}
This yields a monad $ (\_)^{*u}$ on $[T,\Set]$ (\lstinline!opt_monad u!).

For a map $f:V\to W$ in $\TS{T}$ and $w\in W(u)$, we call 
    \[ \defaultmap_u(f,w) : V^{*u} \to W\] 
the coproduct map defined by
 \[ \defaultmap_u (f, w) (x) := \begin{cases}
                                   w,&\text{if } x = * \\
                                   f_t (v),&\text{if } x = v \in V_t.
                                \end{cases}
    \]

Given a monad $P$ over $[T,\Set]$ and a $P$--module $M$ with codomain $[T,\Set]$, we define the \emph{derived module} w.r.t.\ $u\in T$ by setting
\[ 
   M^u (V) := M (V^{*u}). \]
For a morphism $f\in\Hom(V,P(W))$ the module substitution for the derived module is given by
\[ \mkl[{M^u}]{f} := \mkl[M]{{}_uf}. \]
Here the ``shifted'' map \[\shift{f}{u} : V^{*u} \to P (W^{*u})\] 
is defined as %
\[ \shift{f}{u} := \defaultmap\bigl((\comp{f}{Pi}), \we(*)\bigr), \]
the map $i : W\to W^*$ being the inclusion map.
\begin{example}
When $P$ is a monad of terms over free variables, 
the map $\shift{f}{u}$ sends the additional variable of $V^{*u}$ to $\eta^{P}(*u)$, 
i.e.\ to the term consisting of just the ``freshest'' free variable.
When recursively substituting with a map $f : V \to P W$, 
terms under a constructor which binds a variable of type $u$
such as $\lambda_u$ must be substituted using the shifted map $\shift{f}{u}$.
Examples are given in Ex.\ \ref{ex:ulc_monad} for the untyped case and Ex.\ \ref{ex:slc_monad}
for the typed case.
\end{example}



\noindent
Derivation is an endofunctor on the category of $P$--modules with codomain $\TS{T}$.

A constructor can bind several variables at once. 
Given a list \lstinline!l! over \lstinline!T!, the multiple addition of variables with (object language) types according to \lstinline!l! to a set of variables \lstinline!V! is defined by recursion over \lstinline!l!. For this enriched set of variables we introduce the notation \lstinline!V ** l!.
\begin{lstlisting}
Fixpoint pow (l : [T]) (V : ITYPE T) : ITYPE T :=
  match l with
  | nil => V
  | b::bs => pow bs (opt b V)
  end.
\end{lstlisting}
Being a monad, \lstinline!opt! is functorial, as is the multiple addition of variables \lstinline!pow!. 
On morphisms the \lstinline!pow! operation is defined by recursively applying the functoriality of \lstinline!opt!, where for the latter we use a special notation with a prefixed hat.
\begin{lstlisting}
Fixpoint pow_map (l : [T]) V W (f : V ---> W) : 
         V ** l ---> W ** l :=
  match l return V ** l ---> W ** l with
  | nil => f
  | b::bs => pow_map (^f)
  end.
\end{lstlisting}
In the same manner the multiple shifting 
\begin{lstlisting}
Fixpoint lshift (l : [T]) (V W: ITYPE T) (f : V ---> P W) : 
           V ** l ---> P (W ** l) := ...
\end{lstlisting}
is defined.

\vspace{1.5ex}

The pullback operation commutes with products, derivations and fibres:

\begin{lemma} \label{pb_prod}Let $\C$ be a category and $\D$ be a category with products. 
Let $P$ and $Q$ be monads over $\C$ and $\rho : P \to Q$ a monad morphism. Let $M$ and $N$ be $Q$--modules with codomain $\D$. 
Then %
 the following $P$--modules are isomorphic:
 \[ \rho^* (M \times N) \cong \rho^*M \times \rho^*N \enspace .\]
\end{lemma}
\begin{lemma} \label{pb_comm} Consider the setting as in the preceding lemma, with $\C = [T,\Set]$ and $\D = \Set$. 
Let $u$ be an element of $T$.
The following $P$--modules are isomorphic:
\[ \rho^* (M^u) \cong (\rho^*M)^u \]
and 
\[ \rho^* (M_u) \cong (\rho^*M)_u \enspace . \]
\end{lemma}

The carriers of these isomorphisms are families of identity functions, respectively, 
since the carriers of the source and target modules are convertible. As modules, however, source and target are not convertible in \textsf{Coq}.
In our formalization we will have to insert these isomorphisms 
(called \lstinline!PROD_PB!, \lstinline!ITDER_PB! and \lstinline!ITFIB_PB!) in order to make some compositions typecheck.

\section{Signatures \& Representations}\label{section:sig}

An \emph{arity} %
entirely describes the type and binding behaviour of a \emph{constructor}, and a \emph{signature} is a family of arities.
A signature may be seen as an abstract way of storing all relevant information about a syntax.

Given a signature $S$, a \emph{representation of $S$} is given by any monad $P$ (on a specific category) which is equipped 
with some additional structure depending on $S$.
This additional structure is analoguous to the operations $Z : X$ and $S : X\to X$ 
that a representation of the signature $\mathcal{N}$
(cf.\ Sec.\ \ref{subsec:informal_intro}) in a set $X$ comes with.

Representations of $S$ and their morphisms form a category, which, according to our main theorem, 
has an initial object.

\subsection{Arities \& Signatures}

To any constructor of a syntax we associate an \emph{arity}, which is
intuitively an abstract way of storing all necessary (binding and typing) information about the constructor.
A \emph{signature} is a family of arities. 

To any syntax $\Sigma$ we can associate
its signature, which is simply the family of arities associated to the constructors of $\Sigma$.

We start with an example before giving the general definition:
\begin{example}\label{ex:slc_arities}
 Consider Ex.\ \ref{ex:slc_def} of the simply--typed lambda calculus. Given two types
 $s, t\in \mathcal{T}$, the arity associated to the constructor $\App(s,t)$ is 
\[ \app(s,t) := [](s \Rightarrow t), []s \longrightarrow t  \enspace , \]
meaning that $\App(s,t)$ takes two arguments, a term of type $s \Rightarrow t$ and one of 
  type $s$, yielding a term of type $t$. The empty lists signify that in both arguments 
  no variables will be bound.

  The arity associated to the constructor $\Abs(s,t)$ is 
\[ \abs(s,t) :=  [s]t \longrightarrow (s\Rightarrow t) \enspace , \]
  where in the argument one variable of type $s$ is bound by the constructor, yielding a term 
  of arrow type.
\end{example}

\begin{example}
  Untyped syntax may be considered as simply--typed over the singleton set of types, 
  hence falling into the class of languages we consider.
  In that case the only information an arity needs to give about a constructor is its number of arguments
  and the number of variables bound in each argument. The example of the untyped lambda calculus
  (cf.\ Ex.\ \ref{ex:ulc_def}) shows such simplified arities.
\end{example}

For the formal definitions let us fix a set $T$ of object language types.

\begin{definition}
A \emph{$T$--arity} is a family of types consisting of $t_i \in T$ for $i = 0, \ldots, n$ and
$t_{i,j}\in T$ for all $j=1,\ldots,m_i$ and all $i = 1,\ldots, n$, written
\begin{equation} [t_{1,1} \ldots t_{1,m_1}]t_1 , \ldots , [t_{n,1} \ldots t_{n,m_n}]t_n \to t_0 \label{eq:arity}\end{equation}
or shorter
\[(\vec{s}_{1})t_1,\ldots,(\vec{s}_{n}) t_n \to t_0 \]
where $\vec{s}_k$ denotes the list of types $t_{k,1} \ldots t_{k,m_k}$.
A \emph{$T$--signature} is a family of $T$--arities. 
\end{definition}

A signature could be implemented as a pair consisting of a type \lstinline!sig_index! -- which is used for indexing the arities -- and 
a map from the indexing type to the actual arity type, which is simply built using lists -- using a Haskell--like notation -- and products.
\begin{lstlisting}
Record Signature : Type := {
  sig_index : Type;
  sig : sig_index -> [[T] * T] * T }.
\end{lstlisting}

A slight modification however turns out to be useful. During the construction of the initial representation a universal quantification 
over arities with a given target type is needed. We 
choose to define a signature to be a function which maps 
each \lstinline!t : T! to the set of arities whose output type is the given \lstinline!t!. 
In other words, the parameter \lstinline!t! of \lstinline!Signature_t! replaces the second component of the arities.
\begin{lstlisting}
Record Signature_t (t : T) : Type := {
  sig_index : Type ;
  sig : sig_index -> [[T] * T] }.

Definition Signature := forall t, Signature_t t.
\end{lstlisting}

\begin{example} (Impl.\ of Ex.\ \ref{ex:slc_arities}) As an example we discuss the signature of the simply typed lambda calculus.
At first we define an indexing type \lstinline!TLC_index_t! for each object type \lstinline!t : T!. 
After that, we build an indexed signature \lstinline!TLC_sig! mapping each index to its collection of arities.
\begin{lstlisting}
Inductive TLC_index : T -> Type := 
   | TLC_abs : forall s t : T, TLC_index (s --> t)
   | TLC_app : forall s t : T, TLC_index t.

Definition TLC_arguments : forall t, TLC_index t -> [[T] * T] :=
    fun t r => match r with 
      | TLC_abs u v => (u::nil,v)::nil
      | TLC_app u v => (nil,u --> v)::(nil,u)::nil
      end.

Definition TLC_sig t := Build_Signature_t t 
      (@TLC_arguments t).
\end{lstlisting}
\end{example}
The example signature of PCF is given in the \textsf{Coq} source files.

\subsection{Representations}

We summarize the preceding sections using the example of $\LC$:
\begin{itemize}
 \item The map $ V \mapsto \LC(V)$ can be given the structure of a monad 
 $\LC : \Set \to \Set$. 
 \item %
 The constructor $\App : \LC \times \LC \to \LC$ is a morphism of $\LC$--modules,
and so is $\Abs : \LC^* \to \LC$. 
 \item The syntax of $\LC$, i.e.\ the arguments and binding behaviour of its constructors,
is stored entirely in the \emph{signature} $\mathcal{LC}$ of $\LC$.
\end{itemize}

\noindent
\emph{Representations} of $\mathcal{LC}$ are obtained by \emph{abstracting from the monad $\LC$}:
\begin{example}\label{ex:ulc_rep}
  A representation $R$ of the untyped lambda calculus is given by 
  \begin{itemize}
   \item a monad $P$ over the category $\Set$ of sets and
   \item two morphisms of modules
      \[\App^R : P \times P \to P \enspace , \quad \Abs^R : P^* \to P \enspace . \] 
  \end{itemize}
\end{example}

The simply--typed lambda calculus as an example of a typed syntax is treated in Ex.\ \ref{ex:slc_rep}, after the general definitions.

In the general case, given a set $T$ of object types, a $T$--arity $\alpha$ associates to any monad $R$ over the category 
$\TS{T}$ two $R$--modules:
a target module $\cod(\alpha,R)$, which is of the form $R_t$ for some $t \in T$, and a more complex source module $\dom(\alpha,R)$.
The latter module is built from products (when the constructor in question takes more than one argument) and
derivations (for binding of variables) of fibre modules of the form $R_s$.

A representation of the arity $\alpha$ in the monad $R$ is given by a morphism of $R$--modules $\dom(\alpha,R) \to \cod(\alpha,R)$:

\begin{definition}
Let  $\alpha := (\vec{s}_{1})t_1 , \ldots , (\vec{s}_{n}) t_n \to t_0$ be a $T$--arity and $R$ be a monad on $[T,\Set]$. 
A \emph{representation} of the arity $\alpha$ in the monad $R$ is an $R$--module morphism
\begin{equation*}
 r_{\alpha}^{R} \colon (R^{\vec{s}_1})_{t_1} \times \ldots \times (R^{\vec{s}_n})_{t_n} \to R_{t_0} \enspace ,
\end{equation*}
where $R^{\vec{s}}$ is the derivation of $R$ associated to the list $(\vec{s})$ of object types obtained by iterating the derivation endofunctor.
We write $\alpha = \ell \to t_0$ for the above arity and $\prod_{\ell}R$ for the domain module.
\end{definition}

\begin{definition}\label{def:rep}
A \emph{representation $R$ of a $T$--signature} $S$ is given by a monad $P : \TS{T} \to \TS{T}$  and a 
representation of each arity $\alpha$ of $S$ in $P$, that is, a
family of $P$--module morphisms 
\[ \alpha^R : \dom(\alpha, R) \to \cod(\alpha,R) \enspace . \]
  
\end{definition}

\begin{remark}\label{rem:coercion_rep_mod}
 Given a representation $R$, we will denote by $R$ also its underlying monad, i.e.\ we will omit the projection to its first component. 
However, it is possible to define two different representations $R$ and $R'$ of a signature in one and the same monad $P$.
\end{remark}

\begin{example}\label{ex:slc_rep}
A \emph{representation $R$ of $\SLC$} is
any tuple of a monad $P$ over $\TS{\mathcal{T}}$ together with two families %
of $P$--module morphisms 
\[\App(s,t)^R : P_{s\Rightarrow t} \times P_s \to P_t \enspace , \quad \Abs(s,t)^R : P^s_t \to P_{s\Rightarrow t} \enspace , \] 
where $s$ and $t$ range over $\mathcal{T}$.
The reader might want to switch back to Ex.\ \ref{ex:slc_arities} and compare how the source and target modules
of those morphisms of modules are determined by the arities $\app(s,t)$ and $\abs(s,t)$.
\end{example}

\subsection{Morphisms of Representations}\label{sec:mor_of_reps}


In the introductory example, a representation of the signature $\mathcal{N}$ is a set $X$ 
together with some ``representation'' data $Z$ and $S$.
A morphism of representations from $(X,Z,S)$ to $(X',Z',S')$ is defined to be a map $f : X \to X'$
between the sets underlying the representations that is compatible
with the representation data in the sense of Display \eqref{eq:nat_mor}.

Another example of initial algebra, which illustrates a constructor with 2 arguments, 
is the signature defining the types of $\SLC$ from Ex.\ \ref{ex:slc_def},
 \[  \mathcal{T} := \{  (*) \mapsto 0 \enspace , \quad (\Rightarrow) \mapsto 2 \} \enspace . \]
A morphism of representations from $(X,*, \Rightarrow)$ to $(X',*',\Rightarrow')$ is given by a map $f : X \to X'$
such that
\begin{equation} \label{eq:tlc_types_rep}
   f(*) = *' \quad \text{and}\quad
  \begin{xy}
    \xymatrix{  **[l] X\times X \ar[d]_{f\times f} \ar[r]^{\Rightarrow} &  X \ar[d]^{f} \\
                 **[l] X'\times X' \ar[r]_{\Rightarrow'} & X'.
    }
  \end{xy}
\end{equation}

Transferring this definition to the representations defined in Def.\ \ref{def:rep} yields that
a morphism $P \to Q$ of such representations is given by a monad morphism $f : P \to Q$ of the underlying monads
such that $f$ is compatible in some sense with the representation data.

However, the map $f$ is a \emph{monad} morphism, while the representation data is given by \emph{module} morphisms.
How can we plug them together in a way similar to what is done in Diagram \eqref{eq:tlc_types_rep} ? 

From Sec.\ \ref{mod_examples} we recall that $f$ can be considered as a $P$--module morphism $f: P \to f^* Q$.
We may then apply to $f$ the functors fibre, derivation and products of the category of $P$--modules to obtain a $P$--module
morphism that is adapted to the domain and codomain of some arity.
 
Furthermore, the pullback functor $f^*$ --- which impacts the substitution structure, but not the underlying 
 functor and natural transformation, as explained in Remark \ref{rem:about_pullback} ---
 can be used to obtain a $P$--module morphism from a $Q$--module morphism. This will be used to turn 
  the representation module morphisms of $Q$ into $P$--module morphisms.


\begin{definition}
Let $P$ and $Q$ be representations of a $T$--signature $S$. 
A \emph{morphism of representations} $f\colon P \to Q$ is a morphism of monads $f \colon P\to Q$ (on the underlying monads) such that the 
following diagram commutes for any arity $\alpha = (\vec{s}_{1})t_1,\ldots,(\vec{s}_{n})t_n \to t_0$ of $S$:
\[
\begin{xy}
\xymatrix @=4pc{
**[l] \prod\limits_{i=1}^n (P^{\vec{s}_i})_{t_i}\ar[r]^{{\alpha}^P} \ar[d]_{\prod\limits_i (f^{\vec{s}_i})_{t_i}} & P_{t_0} \ar[d]^{f_{t_0}}  \\
**[l] f^* \prod\limits_{i=1}^n (Q^{\vec{s}_i})_{t_i} \ar[r]_{f^*({\alpha}^Q)} & f^*Q_{t_0} \\
}
\end{xy}
\]
\end{definition}
To make sense of this diagram it is necessary to recall the constructions on modules of section \ref{mod_examples}. The diagram lives in the category $\Mod^P_{\Set}$. 
The vertices are obtained from the tautological modules $P$ resp.\ $Q$ over the 
monads $P$ resp.\ $Q$ by applying the derivation, fibre and pullback functors as well as by the use of the product in the category $\Mod^P_{\Set}$.
The vertical morphisms are module morphisms induced by the monad morphism $f$, to which functoriality of derivation, fibre and products are applied. 
Furthermore instances of lemmas \ref{pb_prod} and \ref{pb_comm} are hidden in the lower left corner. 
The lower horizontal morphism makes use of the functoriality of the pullback operation, and in the lower right corner we again use the fact that pullback commutes with fibres. 
Diagram \eqref{eq:diag} (on page \pageref{eq:diag}) shows an expanded version where the mentioned isomorphisms are explicitly inserted.

\begin{table}[hbt]
\begin{equation}\label{eq:diag}
 \begin{xy}
  \xymatrix @=3.5pc{
  \prod\limits_{i=1}^n (P^{\vec{s}_i})_{t_i}\ar[rr]^{{\alpha}^P} \ar[d]_{\prod_i (f^{\vec{s}_i})_{t_i}}& {} & P_{t_0} \ar[dddd]^{f_{t_0}}  \\
  \prod_{i=1}^n ((f^*Q)^{\vec{s}_i})_{t_i} \ar[d]_{\prod_i (\cong)_{t_i}} & {} &{} \\
  \prod_{i=1}^n (f^*(Q^{\vec{s}_i}))_{t_i} \ar[d]_{\prod_i \cong}& {} &{} \\
  \prod_{i=1}^n f^*((Q^{\vec{s}_i})_{t_i}) \ar[d]_{\cong}& {} &{} \\
  f^* \prod_{i=1}^n (Q^{\vec{s}_i})_{t_i} \ar[r]_-{f^* ({\alpha}^Q)} & f^* (Q_{t_0}) \ar[r]_{\cong}& (f^*Q)_{t_0}
}
 \end{xy}
\end{equation}
 \centering Expanded diagram for morphisms of representations
\end{table}

\begin{example}\label{ex:ulc_rep_mor}(Ex.\ \ref{ex:ulc_rep} cont.)
    Given representations $R$ and $S$ of $\mathcal{LC}$, a morphism of representations from $R$ to $S$ is given by 
  a monad morphism $f : R \to S$ such that the following diagrams commute:
  \[
   \begin{xy}
    \xymatrix @C=3.5pc{  **[l]R \times R \ar[r]^{\App^R} \ar[d]_{f \times f} & R \ar[d]^{f} \\ **[l]f^*(S\times S) \ar[r]_{f^*(\App^S)} & f^*S }
   \end{xy} \quad
   \begin{xy}
    \xymatrix @C=3.5pc{  **[l]R^*  \ar[r]^{\Abs^R} \ar[d]_{f^*} & R \ar[d]^{f} \\ **[l]f^*S^* \ar[r]_{f^*(\Abs^S)} & f^*S }
   \end{xy}
  \]
\end{example}

\begin{example}\label{ex:slc_rep_mor}(Ex.\ \ref{ex:slc_rep} cont.)
    Given representations $R$ and $S$ of the simply--typed lambda calculus, a morphism of representations from $R$ to $S$ is given by 
  a monad morphism $f : R \to S$ such that for any two object types $s, t\in \mathcal{T}$ the following diagrams commute:
  \[
   \begin{xy}
    \xymatrix @C=3.5pc{  **[l]R_{s\Rightarrow t} \times R_s \ar[r]^{\App(s,t)^R} \ar[d]_{f_{s\Rightarrow t} \times f_s} & R_t \ar[d]^{f_t} \\ 
                         **[l]f^*(S_{s\Rightarrow t}\times S_s) \ar[r]_{f^*(\App(s,t)^S)} & f^*S_t }
   \end{xy} \quad
   \begin{xy}
    \xymatrix @C=3.5pc{  **[l]R^s_t  \ar[r]^{\Abs(s,t)^R} \ar[d]_{f^s_t} & **[r]R_{s\Rightarrow t} \ar[d]^{f_{s\Rightarrow t}} \\ 
                         **[l]f^*S^s_t \ar[r]_{f^*(\Abs(s,t)^S)} & **[r]f^*S_{s\Rightarrow t} }
   \end{xy}
  \]
\end{example}

In the formalization, the aforementioned isomorphisms would have to be inserted in order for the commutative diagram to typecheck,
since the isomorphic modules are not convertible.
This would result in quite a cumbersome formalization with decreased readability. 

Instead we implement the left vertical morphism from scratch, that is, we define the data of the map first
and prove afterwards that it is indeed a morphism of modules.
This decision entails another design decision: in \textsf{Coq} it is much more convenient to define a map on 
an inductive data type than on a recursively defined one.
It is hence advantageous to also build the domain module from scratch, instead of by applying recursively 
the categorical product of modules.
Given an arity $\alpha = \ell \to t$ and a monad $R$, we define at first the map
$V \mapsto (\prod_{\ell} P) (V)$ and later equip this map with a module substitution verifying the necessary properties.

Given an arity $(\vec{s}_1)t_1, \ldots ,(\vec{s}_n)t_n\to t_0 $ (or shorter $\ell \to t_0$) and a monad $P$, 
we have to construct the module $\prod_{i=1}^n (P^{\vec{s}_i})_{t_i} = \prod_{\ell} P$.
Its carrier, being a kind of heterogeneous list, is given as an inductive type parametrized by a set of variables \lstinline!V! 
and dependent on an arity (resp. its domain component). For the definition of the carrier, we actually do not need 
all the information of a monad $P$, but just its underlying map on objects of the category $\TS{T}$ -- in the code 
given by the section variable \lstinline!M!:
\begin{lstlisting}
Variable M : (ITYPE T) -> (ITYPE T).
Inductive prod_mod_c (V : ITYPE T) : [[T] * T] -> Type :=
  | TTT :  prod_mod_c V nil 
  | CONSTR : forall b bs, 
       M (V ** (fst b)) (snd b) -> prod_mod_c V bs -> prod_mod_c V (b::bs).
\end{lstlisting}
Given now a module \lstinline!M! over some monad \lstinline!P!, 
the module substitution \lstinline!mkleisli:= pm_mkl! for the module carrier \lstinline!prod_mod_c M!
is defined by recursion on this list--like structure, applying the module substitution
\lstinline!mkleisli! of the module \lstinline!M! in each component:
\begin{lstlisting}
Fixpoint pm_mkl l V W (f : V ---> P W)
      (X : prod_mod_c M V l) : prod_mod_c M W l :=
  match X in prod_mod_c _ _ l return prod_mod_c M W l with
  | TTT => TTT M W
  | CONSTR b bs elem elems =>
         CONSTR (mkleisli (Module_struct := M) (lshift f) (snd b) elem)
                  (pm_mkl f elems)
  end.
\end{lstlisting}
Here the (multiple) shifting \lstinline!lshift! is applied to accommodate the derivations in the 
respective component.

After having proved its module properties (by induction on the list--like structure) and 
hence having defined a module \lstinline!prod_mod l! for each \lstinline!l : [[T] * T]!, 
a type of module morphisms is associated to each arity:
\begin{lstlisting}
Definition modhom_from_arity (ar : [[T] * T] * T) : Type :=
  Module_Hom (prod_mod M (fst ar)) (M [(snd ar)]).
\end{lstlisting} 
where \lstinline!M[(s)]! denotes the fibre of the module \lstinline!M! over \lstinline!s!.

Finally a representation of a signature \lstinline!S! over a monad \lstinline!P! is given by a module morphism for each arity. 
Since the set of arities is indexed by the target of the arities, the representation structure is indexed as well:
\begin{lstlisting}
Variable P : Monad (ITYPE T).
Definition Repr_t (t : T) := 
   forall i : sig_index (S t), modhom_from_arity P ((sig i), t).
Definition Repr := forall t, Repr_t t.
\end{lstlisting}
Here the monad \lstinline!P! is actually seen as a module over itself via the coercion \lstinline!Taut_Mod! mentioned earlier.
After abstracting over the monad \lstinline!P!, 
we bundle the data and define a representation as a monad together with a representation structure over this monad
\footnote{
Here an example of \emph{coercion} occurs. The special notation \lstinline!:>! allows us to omit the projection \lstinline!rep_monad! 
when accessing the monad which underlies a given representation \lstinline!R!. 
We can hence also write \lstinline!R x! for the value of the monad of \lstinline!R! on an object \lstinline!x! of the underlying category. 
This coercion is the formal counterpart to the abuse of notation announced in Remark \ref{rem:coercion_rep_mod}.
}:
\begin{lstlisting}
Record Representation := {
  rep_monad :> Monad (ITYPE T);
  repr : Repr rep_monad }.
\end{lstlisting}

As already mentioned, the carrier of the upper left product module is defined as an inductive type. 
This suggests the use of structural recursion for defining the left vertical morphism of the commutative diagram. 
Given a monad morphism $f : P\to Q$, we apply $f$ to every component of 
$\prod_{\ell} P$
:

\begin{lstlisting}
Fixpoint Prod_mor_c (l : [[T] * T]) (V : ITYPE T) (X : prod_mod P l V) : 
                  f* (prod_mod Q l) V :=
  match X in prod_mod_c _ _ l return f* (prod_mod Q l) V with
  | TTT => TTT _ _ 
  | CONSTR b bs elem elems => 
    CONSTR (f _ _ elem) (Prod_mor_c elems)
  end.
\end{lstlisting}
This function is easily proved to be a morphism of $P$--modules 
\[ \text{\lstinline!Prod_mor!}\colon\prod_{\ell} P \to f^* \prod_{\ell} Q \enspace.\]

The isomorphism in the lower right corner however remains in the formalization, appearing as \lstinline!ITPB_FIB!. 
Its underlying family of morphisms, however, is simply a family of identity functions. 
For an arity \lstinline!a! and module morphisms \lstinline!RepP! and \lstinline!RepQ! representing this arity in monads \lstinline!P! and \lstinline!Q! respectively, 
the definition of the commutative diagram 
reads as follows.
\begin{lstlisting}
Definition commute f RepP RepQ : Prop :=
   RepP ;; f [(snd a)] ==
   Prod_mor (fst a) ;; f* RepQ ;; ITPB_FIB f _ _ 
\end{lstlisting}
A morphism of representations \lstinline!P! and \lstinline!Q! of the signature \lstinline!S! is just a monad morphism from \lstinline!P! to \lstinline!Q! 
together with the commutativity property for each \lstinline!t : T! and each arity (index) \lstinline!i! in the indexing set of \lstinline!S t!:
\begin{lstlisting}
Variables P Q : Representation S.
Class Representation_Hom_struct (f : Monad_Hom P Q) :=
  repr_hom_s : forall t (i : sig_index (S t)), commute f (repr P i) (repr Q i).
Record Representation_Hom : Type := {
  repr_hom_c :> Monad_Hom P Q ;
  repr_hom :> Representation_Hom_struct repr_hom_c }.
\end{lstlisting}
Morphisms of representations can be composed: the composition of the underlying monad morphisms as defined in Lst.\ \ref{lst:monad_comp_id} makes the necessary diagram commute
and hence gives a morphism of representations.
Similarly the identity morphism of monads is a morphism of representations.
Two morphisms of representations are said to be equal if their underlying morphism of monads are equal. 
With these definitions the collection of representations of the signature \lstinline!S! and their morphisms form a category:
\begin{lstlisting}
Program Instance REPRESENTATION_struct : 
              Cat_struct (@Representation_Hom _ S) := {
  mor_oid a c := eq_Rep_oid a c ;
  id a := Rep_Id a ;
  comp P Q R f g := Rep_Comp f g }.
\end{lstlisting}

The following theorem is the main result of our work:

\begin{mtheorem}\label{nice_thm}
Let $S$ be a $T$--signature. Then the category $\Rep(S)$ of representations of $S$ has an initial object $\Sigma(S)$.  
\end{mtheorem}
The formal counterpart of this theorem is the instance declaration for the \lstinline!Initial! type class of Lst.\ \ref{lst:initial_rep}.

\begin{remark}
 The monad underlying the initial representation associates to any $V\in \TS{T}$ the set of terms of the syntax 
 associated to $S$ with free variables in $V$. 
The module morphisms of the initial representation are given by the constructors of this syntax.
\end{remark}

A set--theoretic construction of the syntax as well as a proof of the theorem can be found in Zsid\'o's PhD thesis \cite{ju_phd}. 
In a type--theoretic setting such as \textsf{Coq} the syntax can be defined as an inductive type. 
The next section is devoted to the proof of the theorem, i.e.\ the construction of the initial representation.

\section{The Initial Object}\label{section:STS}

The initial object of the category of representations of the signature $S$ is constructed in several steps:
\begin{itemize}
 \item the syntax associated to $S$ as an inductive data type \lstinline!STS!,
  \item definition of a monad structure \lstinline!STS_Monad! on said data type,
  \item construction of the representation structure \lstinline!STSRepr! on \lstinline!STS_Monad!,
 \item for any representation \lstinline!R!, construction of a morphism \lstinline! init R! from \lstinline!STSRepr! to \lstinline!R!,
  \item unicity of \lstinline!init R! for any representation \lstinline!R!.
\end{itemize}

\subsection{The Syntax associated to a Signature}
The first step is to define a map \lstinline!STS : ITYPE T ---> ITYPE T! -- the monad carrier -- mapping each type family $V$ of variables to the type family of 
terms with free variables in $V$. Since objects of \lstinline!ITYPE T! really are just dependent \textsf{Coq} types (cf.\ Sec.\ \ref{sec:coq_impl_cat}),
this map can be implemented as a \textsf{Coq} inductive data type, parametrized by a set of variables and dependent on object types.
Apart from the use of dependent types, the ``data'' parts of this section could indeed be done in any programming language featuring inductive types.

Mutual induction is used, defining at the same time a type \lstinline!STS_list! of heterogeneous lists of terms, 
yielding the arguments to the constructors of $S$. This list type is indexed by arities, such that the constructors can be fed with precisely the 
right kind of arguments.
\begin{lstlisting}
Inductive STS (V : ITYPE T) : ITYPE T :=
  | Var : forall t, V t -> STS V t
  | Build : forall t (i : sig_index (S t)), STS_list V (sig i) -> STS V t
with 
 STS_list (V : ITYPE T) : [[T] * T] -> Type :=
  | TT : STS_list V nil
  | constr : forall b bs, 
      STS (V ** (fst b)) (snd b) -> STS_list V bs -> STS_list V (b::bs).

Scheme STSind := Induction for STS Sort Prop with
       STSlistind := Induction for STS_list Sort Prop.
\end{lstlisting}
The constructor \lstinline!Build! takes 3 arguments: 
\begin{itemize}
 \item an object type \lstinline!t! indicating its output type,
 \item an arity \lstinline!i! (resp.\ its index) from the set of indices with output type \lstinline!t! and
 \item a term of type \lstinline!STS_list V (sig i)! carrying the subterms of the term to construct.
\end{itemize}
Note that \textsf{Coq} typing ensures the correct typing of all constructible terms of \lstinline!STS!, 
a techique called \emph{intrinsic typing}.

The \lstinline!Scheme! command generates a mutual induction scheme for the defined pair of types.

The latter type, \lstinline!STS_list!, is actually isomorphic to the type \lstinline!prod_mod_c STS!. 
This duplication of data could hence have been avoided by 
defining \lstinline!STS! as a \emph{nested} inductive type as follows, instead of using mutual induction.
\begin{lstlisting}
Inductive STS (V : ITYPE T) : ITYPE T :=
  | Var : forall t, V t -> STS V t
  | Build : forall t (i : sig_index (S t)), prod_mod_c STS V (sig i) -> STS V t.
\end{lstlisting}
%
However, we use the mutual inductive version because it allows us to define functions on those types by mutual recursion
rather than nested recursion. 
We found nested recursive functions to be difficult to reason about, whereas the mutual induction principle produced by
the \lstinline!Scheme! command makes reasoning about mutual recursive functions as easy as one could wish, compensating 
for any inconvenience caused by the duplication of data (cf.\ Sec.\ \ref{sec:rep_in_sts}).

\subsection{Monad Structure on Syntax}

We continue by defining a monad structure on the map \lstinline!STS!. Again, due to our choice of implementing sets as \textsf{Coq} types 
(cf.\ Sec.\ \ref{sec:coq_impl_cat}), the maps we need are really just \textsf{Coq} functions.
As in the special case of $\LC$ (cf.\ Ex.\ \ref{ex:ulc_monad}) and $\SLC$ (cf.\ Ex.\ \ref{ex:slc_monad}),
the term--as--variable constructor \lstinline!Var! serves as monadic map $\eta$.
The substitution map \lstinline!subst! is defined using two helper functions \lstinline!rename! (providing functoriality) and \lstinline!_shift!
(serving the same purpose as in Ex.\ \ref{ex:ulc_monad}). Renaming and substitution, being recursive functions on the inductive data types, 
are implemented using mutual recursion:
\begin{lstlisting}
Fixpoint rename V W (f : V ---> W) t (v : STS V t):=
    match v in STS _ t return STS W t with
    | Var t v => Var (f t v)
    | Build t i l => Build (l //-- f)
    end
with
  list_rename V t (l : STS_list V t) W (f : V ---> W) : STS_list W t :=
     match l in STS_list _ t return STS_list W t with
     | TT => TT W
     | constr b bs elem elems =>
             constr (elem //- ( f ^^ (fst b)))
                               (elems //-- f)
     end
where "x //- f" := (rename f x)
and "x //-- f" := (list_rename x f).

(* ... *)

Fixpoint subst (V W : ITYPE T) (f : V ---> STS W) t (v : STS V t) :
  STS W t := match v in STS _ t return STS W t with
    | Var t v => f t v
    | Build t i l => Build (l >>== f)
    end
with
  list_subst V W t (l : STS_list V t) (f : V ---> STS W) : STS_list W t :=
     match l in STS_list _ t return STS_list W t with
     | TT => TT W
     | constr b bs elem elems =>
       constr (elem >== (_lshift f)) (elems >>== f)
     end
where "x >== f" := (subst f x)
and "x >>== f" := (list_subst x f).
\end{lstlisting}
The monadic properties that the substitution should verify, resemble the lemmas one would prove in order to establish ``program correctness''.
As an example, the third monad law reads as
\begin{lstlisting}
Lemma subst_subst V t (v : STS V t) W X (f : V ---> STS W)
             (g : W ---> STS X) :
     v >== f >== g = v >== f;; subst g.
Proof.
  apply (@STSind
    (fun (V : T -> Type) (t : T) (v : STS V t) => forall (W X : T -> Type)
          (f : V ---> STS W) (g : W ---> STS X),
        v >== f >== g = v >== (f;; subst g))
   (fun (V : T -> Type) l (v : STS_list V l) =>
       forall (W X : T -> Type)
          (f : V ---> STS W) (g : W ---> STS X),
        v >>== f >>== g = v >>== (f;; subst g) ));
  t5.
Qed.
\end{lstlisting}
Its proof script is a typical example; most of those lemmas are proved using the induction scheme \lstinline!STSind! 
-- instantiated with suitable properties -- followed by a single custom tactic which finishes off the resulting subgoals, 
mainly by rewriting with previously proved equalities.

After a quite lengthy series of lemmas we obtain that the function \lstinline!subst! and the variable--as--term constructor \lstinline!Var! turn \lstinline!STS! into a monad:
\begin{lstlisting}
Program Instance STS_monad : Monad_struct STS := {
  weta := Var ;
  kleisli := subst }.
\end{lstlisting}

\subsection{A representation in the Syntax}\label{sec:rep_in_sts}

The representational structure on \lstinline!STS! is defined using the \lstinline!Build! constructor. 
For each arity \lstinline!i! in the index set \lstinline!sig_index (S t)! we must give a morphism of modules from \lstinline!prod_mod STS (sig i)! to 
\lstinline!STS [(t)]!. 
Since the constructor \lstinline!Build! takes its argument from \lstinline!STS_list! and not from the isomorphic \lstinline!prod_mod STS!, 
we precompose with one of the isomorphisms between those two types:
\begin{lstlisting}
Program Instance STS_arity_rep (t : T) (i : sig_index (S t)) : Module_Hom_struct 
       (S := prod_mod STS (sig i)) (T := STS [(t)]) 
   (fun V X => Build (STSl_f_pm X)).
\end{lstlisting}
The only property to verify is the compatibility of this map with the module substitution, which we happily leave to \textsf{Coq}.

The result is the object \lstinline!STSRepr! of the category \lstinline!REPRESENTATION S!:
\begin{lstlisting}
Record STSRepr : REPRESENTATION S := Build_Representation (@STSrepr).
\end{lstlisting}

\subsection{Weak Initiality}

In the introduction we gave the equations that a morphism of representations of the natural numbers should verify.
Reading those equations as a rewrite system from left to right yields a way to define iterative functions on the
natural numbers.
This idea is also used in order to define a morphism from \lstinline!STSRepr! to any representation \lstinline!R! of the signature \lstinline!S!:
a term of \lstinline!STS!, whose root is a constructor \lstinline!Build t i! for some object type \lstinline!t! and an arity \lstinline!i!,
is mapped recursively to the image -- of the recursively computed argument -- 
under the corresponding representation \lstinline!repr R i! of \lstinline!R!. 
This definition for a morphism of representations will turn out to be the only one possible, leading to initiality.

Formally, the carrier \lstinline!init! of what will be the initial morphism  from \lstinline!STSRepr! to \lstinline!R! 
is defined as a mutually recursive \textsf{Coq} function:
\begin{lstlisting}
Fixpoint init V t (v : STS V t) : R V t :=
   match v in STS _ t return R V t with
   | Var t v => weta (Monad_struct := R) V t v
   | Build t i X => repr R i V (init_list X)
   end
with
 init_list l (V : ITYPE T) (s : STS_list V l) : prod_mod R l V :=
   match s in STS_list _ l return prod_mod R l V with
   | TT => TTT _ _
   | constr b bs elem elems =>
        CONSTR (init elem) (init_list elems)
   end.
\end{lstlisting}
where the function \lstinline!init_list! applies \lstinline!init! to (heterogeneous) lists of arguments.
We have to show that this function is (a) a morphism of monads and (b) a morphism of representations.

Several lemmas show that \lstinline!init! commutes with renaming/lifting (\lstinline!init_lift!), 
shifting (\lstinline!init_shift!) and substitution (\lstinline!init_kleisli!):
\begin{lstlisting}
Lemma init_lift V t x W (f : V ---> W) : init (x //- f) = lift f t (init x).
Lemma init_shift a V W (f : V ---> STS W) : forall (t : T) (x : opt a V t),
    init (x >>- f) = x >>- (f ;; @init _).
Lemma init_kleisli V t (v : STS V t) W (f : V ---> STS W) :
  init (v >== f) = kleisli (f ;; @init _ ) t (init v).
\end{lstlisting}
 
The latter property is precisely one of the axioms of morphisms of monads (cf.\ Def.\ \ref{def_monad_hom}, rectangular diagram). 
The second monad morphism axiom which states compatibility with the $\eta$s of the monads involved is 
fulfilled by definition of \lstinline!init! -- it is exactly the first branch of the pattern matching. 
We hence have established that \lstinline!init! is (the carrier of) a morphism of monads:
\begin{lstlisting}
Program Instance init_monadic : Monad_Hom_struct (P:=STSM) init.
Record init_mon := Build_Monad_Hom init_monadic.
\end{lstlisting}

Very much less work is then needed to show that \lstinline!init! also is a morphism of representations:
\begin{lstlisting}
Program Instance init_representic : Representation_Hom_struct init_mon.
\end{lstlisting}

\subsection{Uniqueness \& Initiality}

Its uniqueness is expressed by the following lemma:
\begin{lstlisting}
Lemma init_unique : forall f : STSRepr ---> R , f == init_rep.
\end{lstlisting}
Instead of directly proving the lemma, we prove at first an unfolded version which allows to directly apply 
the mutual induction scheme \lstinline!STSind!:
\begin{lstlisting}
Variable f : Representation_Hom STSRepr R.
Hint Rewrite one_way : fin.
Ltac ttt := tt; 
 (try match goal with [t:T, s : STS_list _ _ |-_] => rewrite <- (one_way s);
             let H:=fresh in assert (H:=repr_hom f (t:=t));
             unfold commute in H; simpl in H end);
             repeat (app (mh_weta f) || tinv || tt).

Lemma init_unique_prepa V t (v : STS V t) : f V t v = init v.
Proof.
  apply (@STSind
     (fun V t v => f V t v = init v)
     (fun V l v => Prod_mor f l V (pm_f_STSl v) = init_list v));
  ttt.
Qed.
\end{lstlisting}

Finally we declare an instance of the \lstinline!Initial! type class for the category of representations \lstinline!REPRESENTATION S! with  
\lstinline!STSRepr! as initial object and \lstinline!init_rep R! as the initial morphism towards any other representation \lstinline!R!.
\begin{lstlisting}[label={lst:initial_rep}, caption={Instance of Initial for category of representations}]
Program Instance STS_initial : Initial (REPRESENTATION S) := {
  Init := STSRepr ;
  InitMor R := init_rep R }.
\end{lstlisting}
The proof field \lstinline!InitMorUnique! is filled automatically using the preceding lemma \lstinline!init_unique!.

\section{Conclusions \& Future Work}
We have presented the formalization of a recently proved theorem of representations of typed binding signatures in monads over (families of) sets.
The theorem features the relatively new notion of \emph{module} over a monad and exhibits the structure of constructors as morphisms of modules.

The nature of the theorem made it convenient for computer theorem proving: the proofs are %
straightforward, carrying no %
surprises. Moreover, they are highly technical using (mutual) induction, something our favourite tool
\textsf{Coq} offers good support for.

Some aspects remain unsatisfactory: using type classes and records simultaneously is at least confusing for the reader, 
even if there are good reasons from the implementor's point of view to do so. 
The weak support for nested induction in \textsf{Coq} obliged us to use mutual induction instead, 
leading to some duplication of data and hence another unnecessary source of confusion.

Other aspects, such as the implementation of syntax in an efficient way, 
i.e.\ without any extrinsic typing device, could be solved due to \textsf{Coq}'s good support for dependent types.

The formalization is split into a general library of category theoretic concepts and a theory--specific part 
comprising the formalization of sections \ref{section:sig} and \ref{section:STS}. 
According to \lstinline!coqwc!\footnote{
The tool \lstinline!coqwc!, part of the standard \textsf{Coq} tools, counts the number of lines in a \textsf{Coq} source file, 
classified into the 3 categories \emph{specification}, \emph{proof} and \emph{comment}.
} 
the latter consists of approx.\ 400 lines of specification and 600 lines of proof. 
The proofs are mostly done in a semi--automated way, employing a proof style promoted by Chlipala in his online book \cite{cpdt}, 
as well as in a published user tutorial \cite{jfr:cpdt}.
An earlier version using a more standard proof style included about 900 lines of proof. 
This reduction is mainly due to the fact that proof automation also stimulates reuse of code -- here reuse of proof code -- 
similarly to how polymorphism does for data structures and functions.
However, we do not claim to be experts in proof automation, nor do we have ``one tactic to rule them all''.

The first author is working on extending the presented result by adding different features. 
A first generalization \cite{2011arXiv1107.4751A} is to enlarge the category of representations to allow
for representations of a $T$-signature in a 
monad over $[U,\Set]$ for a given ``translation of object types'' $f : T\to U$. 
In this way translations from one programming language to another --- over different object types --- can be considered as  
initial morphisms in the category of representations of the source language.

This extension yields a difficulty when one attempts to formalize the theorem in \textsf{Coq}:
for such translations of types, say, $f$, $g$ and $h$, (propositional) 
equalities of the form $h(t) = g(f(t))$ arise, as well as equations such as $f (s\Rightarrow t) = f(s) \Rightarrow f(t)$
for a hypothetical type constructor $(\Rightarrow)$. 
Intrinsic typing expresses typing judgements of some language $L$ by type dependency. 
However, even in the presence of a proof of equality $t = s$ of two object types $s$ and $t$, the types
$L(V)(s)$ and $L(V)(t)$ (for a type family of variables $V$) are not convertible.
In order to consider a term $p \in L(V)(s)$ to have type $t$ instead, one 
would need explicit
type casts and, later, their elimination. This would introduce, in the formalization, a difficulty
which does not arise in the informal mathematics.
 Our \textsf{Coq} library contains two different translations from PCF to $\LC$ which illustrate the 
  heavy use of casts.

Secondly, syntax usually comes with a reduction relation, 
which we model by considering sets \emph{equipped with a preorder} \cite{2011arXiv1107.5252A}.
This change is reflected by passing from monads over (families of) sets to \emph{relative} monads
from sets to preorders. 
We introduce \emph{inequations} for the specification of reduction relations. 
A language with reductions is given by 
 a signature $S$, which specifies the terms of the syntax, as well as of a set of inequations $A$ for that syntax.
The category of representations of $(S,A)$ is defined to be the full subcategory of representations of $S$ that
verify all the inequations of $A$.
We prove that this category has an initial object.
The implementation of this theorem is available on the first author's web page \footnote{\url{http://math.unice.fr/~ahrens}}.
%

%

\paragraph{Acknowledgements}
The theorem was implemented in \textsf{Coq} by the first author during a stay at Universit\`a degli Studi di Firenze, Italy, 
financially supported by the Conseil G\'en\'eral des Alpes--Maritimes CG06.

We wish to thank Andr\'e Hirschowitz and Marco Maggesi for many discussions on the subject and
help with \textsf{Coq}.

Furthermore, we are grateful to Assia Mahboubi for letting us use her \textsf{Coq} syntax file for the \texttt{listings} package.

Last but not least we thank the reviewers and the handling editor of JFR for their valuable comments and careful proofreading.

\bibliographystyle{alpha}
\bibliography{literature}

\end{document}